\documentclass[floatfix,showpacs,pra,aps,twocolumn]{revtex4}

 \usepackage{bm}
 \usepackage{amsmath}
 \usepackage{amssymb}
 \usepackage{latexsym}
 \usepackage{amsfonts}
 \usepackage{epsfig}
 \usepackage{psfrag}

 \newcommand{\ket}[1]{\left|#1\right>}
 \newcommand{\bra}[1]{\left<#1\right|}
 \newcommand{\expval}[1]{\left< #1 \right>}
 \newcommand{\opav}[2]{{\left<\left< #1 \right>\right>}_{#2}}
 \newcommand{\braket}[2]
 {\left<#1|#2\right>}
 \newcommand{\nn}{\nonumber\\}
 
 \newcommand{\f}[1]{\mbox{\boldmath$#1$}}

 \newcommand{\bea}{\begin{eqnarray}}
 \newcommand{\ea}{\end{eqnarray}}
 \newcommand{\eea}{\end{eqnarray}}
 \newcommand{\ord}{{\cal O}}
 
 \newcommand{\traceB}[1]{{\rm Tr_B}\left\{ #1 \right\}}
 \newcommand{\abs}[1]{{\left| #1 \right|}}
 
 \newcommand{\HS}{H_{\rm S}}
 \newcommand{\HB}{H_{\rm B}}
 \newcommand{\HI}{H_{\rm SB}}
 \newcommand{\RS}{\rho_{\rm S}}
 \newcommand{\RB}{\rho_{\rm B}}
 \newcommand{\sinc}[1]{{\rm sinc}\left[#1\right]}

\begin{document}

\title{Preservation of Positivity by Dynamical Coarse-Graining}

\author{Gernot Schaller$^*$ and Tobias Brandes}

\affiliation{Institut f\"ur Theoretische Physik, Technische Universit\"at Berlin, Hardenbergstr. 36, 10623 Berlin, Germany}

\begin{abstract}
We compare different quantum Master equations for the time evolution of the reduced density matrix. 
The widely applied secular approximation (rotating wave approximation) 
applied in combination with the Born-Markov approximation 
generates a Lindblad type master equation ensuring 
for completely positive and stable evolution and is typically well applicable for optical baths. For phonon baths however, the 
secular approximation is expected to be invalid. 
The usual Markovian master equation does not generally preserve positivity
of the density matrix. As a solution we propose a coarse-graining approach with a dynamically adapted coarse graining time scale.
For some simple examples we demonstrate that this preserves the accuracy of the integro-differential Born equation.
For large times we analytically show that the secular approximation master equation is recovered.
The method can in principle be extended to systems with a dynamically changing system Hamiltonian,
which is of special interest for adiabatic quantum computation.
We give some numerical examples for the spin-boson model of cases where a spin system thermalizes rapidly, and other examples where thermalization
is not reached.
\end{abstract}

\pacs{
03.67.-a, 
03.67.Lx 
}

\maketitle


\section{Introduction}

The discovery that quantum computers would have much stronger capabilities for solving
certain kinds of problems (such as number factoring \cite{shor1997a} or database search \cite{grover1997a}) 
than their classical counterparts has initiated a lot of research in quantum information theory \cite{nielsen2000}.

Unfortunately, the fragile quantum coherence necessary for the superior performance 
of quantum computers is very sensitive to the inevitable interaction with the environment 
such that theoretical understanding of this process -- called decoherence -- is absolutely necessary \cite{breuer2002}.

For simple models (such as, e.g., a single spin or harmonic oscillator coupled to a thermalized bath of harmonic oscillators)
and for sufficiently complex couplings to the reservoir after some time the system will equilibrate 
in a thermal state with the bath temperature \cite{breuer2002}.
This behavior would be consistent with our classical expectations.

A recent idea is to protect the quantum information by encoding the solution
to a given problem in the (unknown) ground state of a problem Hamiltonian $H_{\rm P}$.
Given sufficient experimental control of the system Hamiltonian and a reservoir at 
sufficiently low temperature $k_{\rm B} T \ll \Delta E_{\rm min}$ 
(where $\Delta E_{\rm min}$ denotes the energy gap above the ground state energy of $H_{\rm P}$), 
the ground state should be robust against decoherence in the sense that decoherence would always 
drive the system towards its ground state.
In principle, one could then prepare the quantum system in any accessible state and wait sufficiently
long until the equilibration has taken place.
Unfortunately, this mere-cooling approach is not expected to be very efficient, 
since the relaxation rate may be very small \cite{khidekel1995a} or (in extreme cases) 
the system might get stuck in a local minimum \cite{childs2001}.

A possible solution was proposed with the concept of adiabatic quantum computation \cite{farhi2001a}:
The system is initially subject to a simple Hamiltonian $H_{\rm I}$ and is prepared in its (easily accessible) ground state.
Then, the Hamiltonian is slowly deformed into the problem Hamiltonian $H_{\rm P}$.
The adiabatic theorem states that if this transformation proceeds slowly enough, 
the quantum state will closely follow the instantaneous ground state \cite{sarandy2004a}.
Finally, for a nearly adiabatic evolution the system state approximates the system ground state to a high degree.
Consequently, the maximum transformation rate (where the final excitations are acceptably small) corresponds
to the computational complexity of the adiabatic algorithm.
For closed systems, it is related to the spectral properties of the time-dependent system Hamiltonian \cite{jansen2007a,schaller2006b}.
For a reservoir at sufficiently low temperatures, this scheme is thought to be robust against decoherence 
\cite{childs2001} and might even be aided by it \cite{amin2008a}.

Unfortunately, the standard framework of deriving master equations relies on some prerequisites that 
are not always fulfilled in realistic systems.
For example, the Markovian approximation widely used is usually only formulated for 
time-independent system Hamiltonians.
In addition, it sometimes leads to master equations that do not preserve positivity of 
the system density matrix \cite{zhao2002a,stenholm2004a}.
Together with trace preservation, positivity grants stability of the density matrix eigenvalues 
and is thus necessary for probability interpretation (cf. \cite{pechukas1994a}) of the density matrix.
Consequently, observables obtained from non-positive density matrices may become unphysical \cite{yu2000a}.
This problem can be cured by the secular approximation. Combined with the
Markovian approximation, it  leads to Lindblad-type \cite{lindblad1976a} master equations
that generically preserve positivity of the density matrix.
Unfortunately, the secular approximation is rather valid for quantum-optical but not for phonon baths \cite{breuer2002}.
Note that there exist non-Markovian master equations that are not of Lindblad type but nevertheless 
preserve positivity by construction. These models however are either phenomenologic \cite{munro1996a,wilkie2000a} 
in the sense that their parameters are not derived from a microscopic Hamiltonian or
they only grant positivity on a restricted set of initial states \cite{tameshtit1996a}.
Especially in view of an experimental optimization of decoherence effects it is, however, necessary to 
relate the parameters in the master equation to those in the microscopic Hamiltonian \cite{spreeuw2007a}.
For example, for a realistic implementation of an adiabatic (or gate-model) quantum computer one would 
expect the qubits to be coupled to phonon degrees of freedom as well, such that a general treatment is advised.
The present article shall present a further step in that direction.

It has been noted \cite{karrlein1997a,vacchini2000a,lidar2001a} that coarse-graining may ensure for positive evolution
of the reduced density matrix. 
However, the coarse-graining timescale so far had to be much larger than the inverse of the bath density of states cutoff. 
In this paper we argue that by adaptively changing the coarse-graining timescale one does not have to obey this
constraint. Beyond this, we show that for infinitely large coarse-graining times we reproduce the 
widely used secular approximation.
We show analytically that for any fixed coarse-graining timescale $\tau \ge 0$, 
the resulting master equations are of Lindblad form, i.e. they preserve positivity and thereby also stability of the density matrix.

This paper is organized as follows:
In section \ref{Sgeneral} we introduce our notation and in section \ref{Smaster} we compare the standard procedure of deriving
quantum master equations from microscopic models with the proposed adaptive coarse-graining scheme.
We make our method explicit by the example of the spin-boson model in section \ref{Sspinboson}.


\section{General Prerequisites}\label{Sgeneral}

We will consider Hamiltonians which can be divided into three parts
\bea\label{Efullham}
H(t) = \HS(t) + \HI + \HB\,,
\eea
where $\HS(t)$ describes the system part, $\HB$ the part acting on the bath 
(with $\left[\HS(t), \HB\right]=0$) and
\bea\label{Eintham}
\HI = \lambda \sum_{\cal A} A_{\cal A} \otimes B_{\cal A}
\eea
denotes the interaction Hamiltonian with the small dimensionless coupling parameter $\lambda\ll 1$
and system operators $A_{\cal A}$ as well as bath operators $B_{\cal A}$ 
(differing coupling constants can be absorbed in the operator definitions). 
Hermiticity is only required for the complete sum  ($\HI = \HI^\dagger$), but by splitting operators
into hermitian and anti-hermitian parts one can always redefine them such that 
\bea
A_{\cal A} = A_{\cal A}^\dagger\,,\qquad
B_{\cal A} = B_{\cal A}^\dagger\,,
\eea
which will be assumed further-on.

The density matrix of the complete system is thought to evolve according to ($\hbar=1$ throughout)
the von-Neumann equation of motion
\bea\label{Evnsch}
\dot \rho (t) = -i \left[\HS(t) + \HI + \HB,\rho(t)\right]\,.
\eea

Denoting the time evolution operators of system and reservoir by 
$U(t)$ and $V(t)$, respectively, we can switch to the interaction picture
\bea
\f{\rho}(t) &=& U^\dagger(t) V^\dagger(t) \rho(t) V(t) U(t)\,,\nn
\f{\HI}(t) &=& U^\dagger(t) V^\dagger(t) \HI V(t) U(t)\nn
&=& \lambda \sum_{\cal A} \left[U^\dagger(t) A_{\cal A} U(t)\right]\otimes
\left[V^\dagger(t) B_{\cal A} V(t)\right]\nn
&\equiv& \lambda \sum_{\cal A} \f{A_{\cal A}}(t) \otimes \f{B_{\cal A}}(t)\,.
\eea
We will denote all operators in the interaction picture by bold symbols throughout.
In the interaction picture, the equation of motion for the density operator (\ref{Evnsch})
transforms into
\bea\label{Evnint}
\f{\dot \rho}(t) = -i \left[\f{\HI}(t),\f{\rho}(t)\right]\,,
\eea
where one can exploit the smallness of the coupling $\lambda$ to apply perturbation theory.

Equation (\ref{Evnint}) can be formally integrated to yield 
\mbox{$\f{\rho}(t) = \f{\rho}(0) -i \int\limits_0^t \left[\f{\HI}(t'), \f{\rho}(t')\right] dt'$}
and re-inserting this result in Eqn. (\ref{Evnint}) one obtains the following exact equation
\bea\label{Erhoexact}
\dot{\f{\rho}}(t) &=& -i \left[\f{\HI}(t), \f{\rho}(0)\right]\nn
&&- \int\limits_0^t \left[\f{\HI}(t), \left[\f{\HI}(t'), \f{\rho}(t')\right]\right] dt'
\eea
for the density operator.


\section{Quantum Master Equations}\label{Smaster}


We will first state the results of the Born-Markov approximation 
without secular approximation (in subsection \ref{SSmarkov}) and with the secular approximation 
(subsection \ref{SSsecular}) in our notation. Afterwards, we will consider the coarse-graining approach in
subsection \ref{SScoarse_graining}.

With the usual assumptions (compare Appendix \ref{Abaneglect}) involving initial factorization of 
the density matrix and neglecting any change in the reservoir part of the density matrix 
in (\ref{Erhoexact}) we obtain the {\bf Born} equation
\bea\label{Eborn}
\f{\dot{\RS}} &=& -i \traceB{\left[\f{\HI}(t), \f{\RS}(0) \f{\RB^0}\right]}\nn
&&-\int\limits_0^t  \traceB{\left[\f{\HI}(t), \left[\f{\HI}(t'), \f{\RS}(t') \f{\RB^0}\right]\right]} dt'\nn
&&+ \ord\{\lambda^3\}\,,
\eea
where $\traceB{\cdot}$ denotes the trace over the reservoir degrees of freedom.
Evaluating the traces leads to the definition of the reservoir correlation functions 
\bea\label{Ecorrfunc}
C_{\cal AB}(\tau) &\equiv& \traceB{e^{+i \HB \tau} B_{\cal A} e^{-i \HB \tau} B_{\cal B} \RB^0}\nn
&=& C_{\cal BA}^*(-\tau)\,,
\eea
and we obtain with $\expval{B_{\cal A}}=0$ (which can always be achieved by a suitable transformation \cite{trafo})
\bea\label{Eborn1}
\dot{\f{\RS}} &=& \lambda^2 \sum_{\cal AB} \int\limits_0^t \Big\{
\left[ \f{A_{\cal B}}(t') \f{\RS}(t'), \f{A_{\cal A}}(t)\right] C_{\cal AB}(t-t')\nn
&&+{\rm h.c.} \Big\}\, dt' + \ord\{\lambda^3\}\,,
\eea
where h.c. denotes the hermitian conjugate.
The integro-differential character of above equation complicates its solution, since analytical solutions
are only possible in very simple cases \cite{kleinekathoefer2004a,schroeder2007a}
(see also subsection \ref{SSexpdecay}), 
and numerical solutions are hampered by the fact that the complete history of $\f{\RS}(t')$ has to be stored 
in order to evolve $\f{\RS}(t)$.


\subsection{Markovian approximation scheme}\label{SSmarkov}

In the usual \cite{breuer2002} Markovian approximation (see Appendix \ref{Amarkov}) one obtains
for constant system Hamiltonians ($\HS(t)=\HS$)
with the half-sided Fourier transforms 
\bea\label{Eft_half}
\Gamma_{\cal AB}(\omega) &\equiv& \int\limits_0^\infty C_{\cal AB}(\tau) e^{i\omega \tau} d\tau
\eea
of the reservoir correlation functions (\ref{Ecorrfunc})
the time-local {\bf Born-Markov} (BM) master equation (here given in the Schr\"odinger picture)
\bea\label{Emaster_mknrwa}
\dot{\RS} &=& -i \left[\HS, \RS(t)\right]\nn
&&+ \lambda^2 \sum_{abcd} \sum_{\cal AB} \Big\{ \Gamma_{\cal AB}(E_b - E_a)\times\nn
&&\times \bra{a}A_{\cal B}\ket{b} \bra{c} A_{\cal A} \ket{d}^* 
\left[\Big(\ket{a}\bra{b}\Big) \RS(t), \Big(\ket{c} \bra{d}\Big)^\dagger\right]\nn
&&+ {\rm h.c.}\Big\}\,,
\eea
where $\HS \ket{a} = E_a \ket{a}$ denote the orthonormal energy eigenbasis.
By construction, Eqn. (\ref{Emaster_mknrwa}) does
preserve trace and hermiticity of $\RS$. 
Note however, that positivity of its solution $\RS(t)$ is not generally preserved \cite{zhao2002a,whitney2008a}, 
see subsection \ref{SSpositive} for some counterexamples.


\subsection{The secular approximation}\label{SSsecular}

In order to restore preservation of positivity in the BM approximation in general, it
is necessary to perform the secular approximation (see Appendix \ref{Asecular}).
Typically, this approximation is known to be well-satisfied for quantum-optical systems, where
it is also known as rotating wave approximation \cite{scully2002}.
In this case, one can combine 
\bea\label{Eevenoddft}
\gamma_{\cal AB}(\omega) &=& \Gamma_{\cal AB}(\omega) + \Gamma_{\cal BA}^*(\omega) = \int\limits_{-\infty}^{+\infty} C_{\cal AB}(\tau) e^{i\omega \tau} d\tau\,,\nn
\sigma_{\cal AB}(\omega) &=& \Gamma_{\cal AB}(\omega) - \Gamma_{\cal BA}^*(\omega)\nn
&=& \int\limits_{-\infty}^{+\infty} C_{\cal AB}(\tau) {\rm sgn}(\tau) e^{i\omega \tau} d\tau
\eea
which yields the {\bf Born-Markov-Secular} (BMS) approximation (in the Schr\"odinger picture)
\bea\label{Elindblad_rwa1}
\dot{\RS} &=& -i \left[\HS, \RS(t)\right]\nn
&&-i \left[\sum_{ab} \tilde \sigma_{ab} \ket{a}\bra{b}, \RS(t)\right]\nn
&&+\sum_{abcd} \tilde \gamma_{ab, cd} \Big[\Big(\ket{a}\bra{b}\Big) \RS(t) \Big(\ket{c}\bra{d}\Big)^\dagger\nn
&&-\frac{1}{2}\left\{\Big(\ket{c}\bra{d}\Big)^\dagger \Big(\ket{a}\bra{b}\Big), \RS(t)\right\} \Big]\,,\nn
\tilde \sigma_{ab} &=& \frac{\lambda^2}{2i} \sum_c \sum_{\cal AB} \sigma_{\cal AB}(E_a-E_c) \delta_{E_b, E_a}\times\nn
&&\times \bra{c}A_{\cal A}\ket{a}^* \bra{c} A_{\cal B} \ket{b}\,,\nn
\tilde \gamma_{ab, cd} &=& \lambda^2 \sum_{\cal AB} \gamma_{\cal AB}(E_b-E_a) \delta_{E_d-E_c, E_b-E_a}\times\nn
&&\times \bra{a}A_{\cal B}\ket{b} \bra{c} A_{\cal A} \ket{d}^*\,.
\eea
Eqn. (\ref{Elindblad_rwa1}) 
has many favorable properties (compare e.g. chapter 3.3 in \cite{breuer2002}):

By construction, it preserves trace and hermiticity of the system density matrix $\RS$.

Since it is of Lindblad \cite{lindblad1976a} form (the matrix $\gamma_{\cal AB}(\omega)$ is positive semidefinite), 
it preserves positivity of the density matrix.

For a thermalized reservoir characterized by the inverse temperature $\beta$, the Fourier transforms 
of the bath correlation functions can be
used to show that the system thermal equilibrium state with the same temperature
\bea\label{Egibbs}
\RS^{\rm th} = \frac{e^{-\beta \HS}}{{\rm Tr_S}\left\{e^{-\beta \HS}\right\}}
\eea
is a stationary state.

If the spectrum of the system Hamiltonian $\HS$ is non-degenerate (implying that $\delta_{E_a, E_b}=\delta_{ab}$), the equations for 
the diagonal elements of $\RS$ in the eigenbasis of $\HS$ completely decouple from the equations for the off-diagonals, and one 
obtains the same transition rates between the populations as with Fermis Golden Rule.


\subsection{Coarse-graining approach}\label{SScoarse_graining}

Eqn. (\ref{Evnint}) is formally solved by
\mbox{$\f{\rho}(t_2) = \f{W}(t_2, t_1) \f{\rho}(t_1) \f{W}^\dagger(t_2, t_1)$}
with the interaction picture time evolution operator
\bea\label{Etimoint}
\f{W}(t_2, t_1) = {\cal T} \exp\left\{-i \int\limits_{t_1}^{t_2} \f{\HI}(t') dt'\right\}\,,
\eea
where the time-dependence of $\f{\HI}(t)$ necessitates the time-ordering \cite{scully2002}
${\cal T} \f{\HI}(t_1) \f{\HI}(t_2) \equiv \f{\HI}(t_1) \f{\HI}(t_2) \Theta(t_1-t_2)
+ \f{\HI}(t_2) \f{\HI}(t_1) \Theta(t_2 - t_1)$,
with $\Theta(x)$ denoting the Heavyside step function.
Now, by expanding $\f{W}(t+\tau, t)$ up to second order in $\lambda$ 
we obtain a second-order approximation to the 
full density matrix
\begin{widetext}
\bea\label{Efullmap}
\f{\rho}(t+\tau) &=& \f{\rho}(t) -i \left[ \int\limits_t^{t+\tau} \f{\HI}(t_1) dt_1, \f{\rho}(t)\right]
-i \left[ \frac{1}{2i} \int\limits_t^{t+\tau} \int\limits_t^{t+\tau} 
\left[\f{\HI}(t_2), \f{\HI}(t_1)\right] \Theta(t_2-t_1) dt_1 dt_2, \f{\rho}(t)\right]\nn
&&+ \int\limits_t^{t+\tau} \int\limits_t^{t+\tau} 
\left[
\f{\HI}(t_1) \f{\rho}(t) \f{\HI}(t_2) - \frac{1}{2}\left\{ \f{\HI}(t_2) \f{\HI}(t_1), \f{\rho}(t)\right\}\right] dt_1 dt_2
+ \ord\left\{\lambda^3\right\}\,.
\eea
\end{widetext}
Since such a truncated finite order approximation to the time evolution operator $\f{W}(t+\tau)$ is still unitary, 
the above map (\ref{Efullmap}) preserves hermiticity, trace and positivity of the full density matrix $\f{\rho}$.
An equivalent expression 
can be obtained from iterative solution of 
equation (\ref{Erhoexact}) by keeping only terms up to $\ord\{\lambda^2\}$.
We will proceed with this second order approximation and 
derive from this fully unitary map 
a non-unitary map for the system part of the density matrix that preserves positivity.

If we neglect the back-action of the system on the bath and assume factorization (Born-approximation) 
as described in Appendix \ref{Abaneglect} we can perform the partial trace over the reservoir degrees of freedom.
By inserting the definition (\ref{Eintham}) in Eqn. (\ref{Efullmap}) and employing the definition of
the reservoir correlation functions (\ref{Ecorrfunc}) we obtain (again working in a frame where $\expval{B_{\cal A}}=0$ \cite{trafo})
\bea\label{Eredmap}
\f{\RS}(t+\tau) &=& \f{\RS}(t)\nn
&&-i \frac{\lambda^2}{2i} \sum_{\cal AB} \int\limits_t^{t+\tau} \int\limits_t^{t+\tau} C_{\cal AB}(t_2-t_1) {\rm sgn}(t_2-t_1)\times\nn
&&\times \left[\f{A_{\cal A}}(t_2) \f{A_{\cal B}}(t_1), \f{\RS}(t)\right] dt_1 dt_2\nn
&&+\lambda^2  \sum_{\cal AB} \int\limits_t^{t+\tau}\int\limits_t^{t+\tau} C_{\cal AB}(t_2-t_1)\times\nn
&&\times\Big[\f{A_{\cal B}}(t_1) \f{\RS}(t) \f{A_{\cal A}}(t_2)\nn
&&-\frac{1}{2} \left\{ \f{A_{\cal A}}(t_2) \f{A_{\cal B}}(t_1), \f{\RS}(t)\right\}
\Big] dt_1 dt_2 + \ord\{\lambda^3\}\nn
&\equiv& \f{\RS}(t) + \tau \f{L^\tau}(t) \f{\RS}(t) + \ord\{\lambda^3\}\,,
\eea
which defines the action of the Liouvillian on the reduced density matrix in the interaction picture.
If one re-arranges the matrix elements of the density matrix as a $N^2$-dimensional vector, the Liouvillian super-operator 
can be understood as an $N^2\times N^2$ (generally non-hermitian) matrix acting on $\f{\RS}$.
In the interaction picture the Liouvillian is small, i.e., since $\expval{B_{\cal A}}=0$ we even have $\f{L^\tau}=\ord\{\lambda^2\}$.

Unfortunately, the above map $\f{\RS}(t)\to\f{\RS}(t+\tau)$ has some shortcomings:
It does not generally preserve positivity of the reduced density matrix $\f{\RS}$
and in addition, if one applies above equation recursively $n$-times with small timesteps $\Delta \tau$ such
that $n \Delta \tau = \tau$ and compares the result with a single iteration of (\ref{Eredmap}), the
difference between the two solutions is much larger than $\ord\{\lambda^3\}$, i.e., the solution
depends on the choice of the stepsize.

By defining time-averaging of an operator ${\cal\hat O}(t)$ over a time interval $[t, t+\tau]$ as
\bea\label{Etimeav}
\opav{\cal \hat O}{[t,t+\tau]} \equiv \frac{1}{\tau} \int\limits_t^{t+\tau} {\cal \hat O}(t') dt'\,,
\eea
we can write Eqn. (\ref{Eredmap}) as 
\bea\label{Emaster_00}
\opav{\f{\dot \RS}}{[t,t+\tau]} &=& \frac{\f{\RS}(t+\tau) - \f{\RS}(t)}{\tau}\nn
&=& \f{L^\tau}(t)\f{\RS}(t) + \ord\{\lambda^3\}\,.
\eea


\subsubsection{Explicit Liouvillian}

It is convenient to insert the even and odd Fourier transforms from (\ref{Eevenoddft}) of the reservoir correlation functions
\bea\label{Efourier}
C_{\cal AB}(\tau) &=& \frac{1}{2\pi} \int\limits_{-\infty}^{+\infty} \gamma_{\cal AB}(\omega) e^{-i \omega \tau} d\omega\,,\nn
C_{\cal AB}(\tau) {\rm sgn}(\tau) &=& \frac{1}{2\pi} \int\limits_{-\infty}^{+\infty} \sigma_{\cal AB}(\omega) e^{-i \omega \tau} d\omega\,,
\eea
and to expand the system operators in the interaction picture in the orthonormal energy eigenbasis 
\mbox{$\HS \ket{a} = E_a \ket{a}$} of the system Hamiltonian (here assumed to be constant)
\bea\label{Econstbas}
\f{A_{\cal A}}(t_2) &=& \sum_{ab} \bra{a} A_{\cal A} \ket{b} e^{i(E_a-E_b)t_2} \ket{a}\bra{b}\,,\nn
\f{A_{\cal B}}(t_1) &=& \sum_{cd} \bra{c} A_{\cal B} \ket{d} e^{i(E_c-E_d)t_1} \ket{c}\bra{d}\,.
\eea
Then, we can use the relation 
\mbox{$\int\limits_t^{t+\tau} e^{i \alpha t'} dt' = \tau e^{i\alpha t} e^{i\alpha \tau/2}\sinc{\frac{\alpha\tau}{2}}$}
with 
\mbox{$\sinc{x}\equiv\frac{\sin(x)}{x}$} together with Eqn. (\ref{Efourier}) and Eqn. (\ref{Econstbas}) in order to make the Liouvillian in Eqn. (\ref{Emaster_00})
explicit. With denoting the energy differences of the system Hamiltonian by $\Omega_{ab}\equiv E_a - E_b$ we arrive at
\begin{widetext}
\bea\label{Eliouvillian}
\f{L^\tau}(t) \f{\RS}(t) &=& -i \left[\sum_{ab} e^{i\Omega_{ab} t} \sigma_{ab}(\tau) \ket{a}\bra{b}, \f{\RS}(t)\right]\nn
&&+\sum_{abcd} e^{i(\Omega_{cd} - \Omega_{ab})t} \gamma_{cd, ba}(\tau) 
\left[ \Big(\ket{c}\bra{d}\Big) \f{\RS}(t) \Big(\ket{b}\bra{a}\Big)^\dagger
-\frac{1}{2}\left\{ \Big(\ket{b}\bra{a}\Big)^\dagger \Big(\ket{c}\bra{d}\Big), \f{\RS}(t)\right\}\right]\,,\nn
\sigma_{ab}(\tau) &=& \frac{\lambda^2 \tau}{4\pi i} e^{i\Omega_{ab}\tau/2} \sum_c \sum_{\cal AB} \bra{c}A_{\cal A}\ket{a}^* \bra{c}A_{\cal B}\ket{b}
\int\limits_{-\infty}^{+\infty} \sigma_{\cal AB}(\omega) \sinc{(w+\Omega_{ca})\frac{\tau}{2}} \sinc{(w+\Omega_{cb})\frac{\tau}{2}} d\omega\,,\nn
\gamma_{cd, ba}(\tau) &=& \frac{\lambda^2 \tau}{2\pi} e^{i(\Omega_{cd}-\Omega_{ba})\tau/2} \sum_{\cal AB}\bra{b}A_{\cal A}\ket{a}^* \bra{c}A_{\cal B}\ket{d}
\int\limits_{-\infty}^{+\infty} \gamma_{\cal AB}(\omega)\sinc{(w+\Omega_{ba})\frac{\tau}{2}}\sinc{(w+\Omega_{cd})\frac{\tau}{2}} d\omega\,.
\eea
\end{widetext}

The coarse-grained derivative on the l. h. s. of Eqn. (\ref{Emaster_00}) generates a time evolution that can be compared with its usual, first order differential
equation counterpart (we denote the corresponding density operators by an overbar and the continuous index $\tau$ referring to the Liouvillian chosen)
\bea\label{Emaster_replacement}
\frac{d}{dt} \f{\bar{\RS^\tau}}(t) &=& \f{L^\tau}(t)\f{\bar{\RS^\tau}}(t)\,.
\eea
If we initialize this differential equation with a known density matrix $\f{\RS^0}$ and evaluate its formal solution at time $t=\tau$ 
(i.e., after exactly the coarse-graining timescale),  we find to first order in $\f{L^\tau}$
\bea
\f{\bar{\RS^\tau}}(\tau) 
&\approx& \left[\f{1}+ \int\limits_0^{\tau} \f{L^\tau}(t') dt' \right] \f{\RS^0}\,.
\eea
This means that when considering the same initial condition, the difference between the 
two time evolutions resulting from Eqns. (\ref{Emaster_replacement}) and (\ref{Emaster_00}) is given by
\bea\label{Edifference}
\f{\bar{\RS^\tau}}(\tau) - \f{\RS}(\tau) &=& \eta(\tau) + \ord\{\left(\f{L^\tau}\right)^2\}\,,\nn
\eta(\tau) &=& \left[ \int\limits_0^{\tau} \f{L^\tau}(t') dt' - \tau \f{L^\tau}(0)\right] \f{\RS^0}\,,
\eea
i.e., essentially by the difference between the coarse-graining-averaged Liouvillian and its initial value.

It follows from Eqn. (\ref{Eliouvillian}) that especially for short coarse 
graining times $\tau \ll \frac{1}{\abs{\Delta E_{\rm max}}}$
(with $\Delta E_{\rm max}$ denoting the maximum energy difference of $\HS$)
the difference will be negligible $\eta(\tau) \approx 0$,
such that the two solutions $\f{\bar{\RS^\tau}}(\tau)$ and $\f{\RS}(\tau)$ are equivalent.
Since we have so far not made any assumption on separating timescales, this also implies that non-Markovian
effects (that are expected for short times where the Markovian approximation does not hold) are within reach 
of the adaptive coarse-graining approach, where the coarse-graining time is chosen to match the physical time.

In the limit of very large coarse-graining times we also obtain $\lim\limits_{\tau\to\infty} \eta(\tau) \approx \f{0}$, 
since then the $\sinc{\cdot}$ functions in (\ref{Eliouvillian}) begin to act like delta functions, see below. 

Also for intermediate coarse-graining times it is easy to see from the structure of the Liouvillian (\ref{Eliouvillian}) 
that the difference $\eta(\tau)$ is bounded throughout.

From now on, we will omit the overbar and write $\RS^\tau(t)$ for the solutions of Eqn. (\ref{Emaster_replacement}).


\subsubsection{The infinite-$\tau$ limit}\label{SSinfinite}

In the limit $\tau\to\infty$ one should note that for discrete $a,b$ and under an integral over $d\omega$ 
with another integrable function 
one has in a distributive sense (see Appendix \ref{Asincidentity})
\bea\label{Esincidentity}
f(\omega, a, b) &\equiv& \lim_{\tau\to\infty} \tau \; \sinc{(w+a)\frac{\tau}{2}} \sinc{(w+b)\frac{\tau}{2}}\nn
&\asymp&2 \pi \delta_{ab} \delta(w+a)\,,
\eea
where $\delta_{ab}$ is a Kronecker symbol and $\delta(w+a)$ denotes the Dirac Delta distribution.
Therefore, we obtain for the Liouvillian matrix elements (\ref{Eliouvillian}) in this limit
\bea
\sigma_{ab}^\infty &=& \frac{\lambda^2}{2 i} \delta_{E_a, E_b} \sum_c \sum_{\cal AB}
\bra{c}A_{\cal A}\ket{a}^* \bra{c}A_{\cal B}\ket{b}\times\nn
&&\times \sigma_{\cal AB}(E_a - E_c)\,,\nn
\gamma_{cd, ba}^\infty &=& \lambda^2 \delta_{E_b-E_a, E_c-E_d} \sum_{\cal AB}\bra{b}A_{\cal A}\ket{a}^*\times\nn
&&\times \bra{c}A_{\cal B}\ket{d} \gamma_{\cal AB}(E_a-E_b)\,,
\eea
and we recover 
the secular approximation (\ref{Elindblad_rwa1})! 
Naturally, this also implies that in the large-time limit, the solution $\f{\RS^{\tau=t}}(t)$ captures
all the favorable properties of Eqn. (\ref{Elindblad_rwa1}). 


\subsubsection{Positivity}

The most striking advantage of the coarse-graining procedure is that not only for very large, 
but for any fixed coarse-graining time $\tau \ge 0$, the resulting first-order differential equations (\ref{Emaster_replacement}) 
are all of Lindblad form \cite{lindblad1976a} and thus intrinsically preserve the positivity of the density matrix.
This is easily seen by switching back to the Schr\"odinger picture, where the time-dependent phases in
(\ref{Eliouvillian}) cancel
\bea\label{Emaster_01}
\dot{\RS^\tau}(t) &\equiv& L^\tau \RS^\tau(t)\nn
&=& -i \left[\HS, \RS^\tau(t)\right]\nn
&&-i \left[\sum_{ab} \sigma_{ab}(\tau) \ket{a}\bra{b}, \RS^\tau(t)\right]\nn
&&+\sum_{abcd} \gamma_{cd, ba}(\tau) \Bigg[ \Big(\ket{c}\bra{d}\Big) \RS^\tau(t) \Big(\ket{b}\bra{a}\Big)^\dagger\nn
&&-\frac{1}{2}\left\{\Big(\ket{b}\bra{a}\Big)^\dagger \Big(\ket{c}\bra{d}\Big), \RS^\tau(t)\right\}\Bigg]\,,
\eea
which implies that the Schr\"odinger picture Liouvillian $L^{\tau}$ only depends on the coarse-graining timescale $\tau$.

First of all, the first two commutator terms correspond to commutators with a hermitian operator, i.e., 
the second term accounts for the unitary action of decoherence (sometimes called Lamb-shift \cite{breuer2002}).
Note that hermiticity of the corresponding effective Hamiltonian follows from the definition of the odd 
Fourier transform versus the half-sided Fourier transform (\ref{Eevenoddft}). 

In order to show that (\ref{Emaster_01}) is a Lindblad form, it remains to be shown that the 
matrix $\gamma_{cd, ba}(\tau)$ is positive semidefinite.
In order to see this we introduce the double indices
$i=(cd)$ and $j=(ba)$ running from 1 to $N^2$ for an $N$-dimensional system Hilbert space.
Then, we can use the short-hand notation 
\mbox{$w_i=E_c-E_d$}, \mbox{$A_{\cal B}^i=\bra{c}A_{\cal B}\ket{d}$}, \mbox{$w_j=E_b-E_a$}, \mbox{$A_{\cal A}^j = \bra{b}A_{\cal A}\ket{a}$} and consider for arbitrary complex-valued 
numbers $x_j$ in Eqn. (\ref{Eliouvillian})
\bea
\sum_{ij} x_i^* \gamma_{ij} x_j &=& \frac{\tau}{2\pi} \int\limits_{-\infty}^{+\infty} \sum_{\cal AB} z_{\cal A}^*(\omega) \gamma_{\cal AB}(\omega) z_{\cal B}(\omega) d\omega\,,
\eea
where we have
\bea
z_{\cal A} &\equiv& \sum_j A_{\cal A}^j \sinc{(w+w_j)\tau/2} e^{i\omega_j \tau/2} x_j\,.
\eea
Now, since the Fourier-transform matrix $\gamma_{\cal AB}(\omega)$ is positive semidefinite (compare also appendix \ref{Asecular})
one has $z_{\cal A}^*(\omega) \gamma_{\cal AB}(\omega) z_{\cal B}(\omega)\ge 0$.
The integral over a strictly positive function can only yield positive results, and since also $\tau \ge 0$
it follows that $\gamma_{ij}(\tau) = \gamma_{cd, ba}(\tau)$ is a positive semidefinite matrix.

This also implies that solutions of the form $\RS^t(t)$ are always positive density matrices, since they correspond
to an interpolation along Lindblad density matrices.


\subsection{Time-dependent Generalization}

Within the coarse-graining approach, it was in Eqn. (\ref{Econstbas}) where it was used for the first time that
the system Hamiltonian $\HS$ was time-independent and had a discrete spectrum. 
Here we will show that coarse-graining generally leads to a time-inhomogeneous (i.e., one with time-dependent 
operators) Lindblad form of master equations and thus always preserves positivity of the density matrix.
With introducing the notation
\bea
\f{\tilde A_{\cal A}}(t, \omega) \equiv \f{A_{\cal A}}(t) e^{i \omega t}
\eea
we can use equations (\ref{Efourier}) and (\ref{Etimeav}) in Eqn. (\ref{Eredmap}) to obtain
\begin{widetext}
\bea
\opav{\f{\dot\RS}}{[t,t+\tau]} &=& 
-i \left[ \frac{\lambda^2 \tau }{4\pi i} \int\limits_{-\infty}^{+\infty} \sum_{\cal AB} \sigma_{\cal AB}(\omega) 
\opav{\f{\tilde A_{\cal A}}(\omega)}{[t,t+\tau]}^\dagger \opav{\f{\tilde A_{\cal B}}(\omega)}{[t,t+\tau]} d\omega, \f{\RS}(t)\right]\nn
&&+\frac{\lambda^2\tau}{2\pi} \int\limits_{-\infty}^{+\infty} \sum_{\cal AB} \gamma_{\cal AB}(\omega) 
\left[ \opav{\f{\tilde A_{\cal B}}(\omega)}{[t,t+\tau]} \f{\RS}(t) \opav{\f{\tilde A_{\cal A}}(\omega)}{[t,t+\tau]}^\dagger\right.\nn
&&\left.- \frac{1}{2} \left\{\opav{\f{\tilde A_{\cal A}}(\omega)}{[t,t+\tau]}^\dagger \opav{\f{\tilde A_{\cal B}}(\omega)}{[t,t+\tau]}, \f{\RS}(t)\right\}
\right]\,.
\eea
\end{widetext}
With the replacements \mbox{$\opav{\f{\dot\RS}}{[t,t+\tau]} \to \f{\dot\RS^{\tau}}$} and 
\mbox{$\f{\RS(t)} \to \f{\RS^{\tau}}(t)$} this becomes a time-inhomogeneous Lindblad 
master equation, since the positivity of the $\gamma_{\cal AB}(\omega)$-matrix at every $\omega$ is guaranteed for reservoirs in thermal equilibrium.
Intuitively, the time-dependence of the Lindblad operators does not destroy positivity, since at any fixed time $t$ one may approximate the time-dependent
operators by a constant-operator Lindblad form, see Appendix \ref{Apositivity} for a more explicit discussion. 
Since the case of slowly varying system Hamiltonians is of special interest in the context of adiabatic quantum computation \cite{childs2001}, 
we outline in Appendix \ref{Aadiabatic} how one could in principle calculate the time-averaged operators
\bea\label{Eoptimeav}
\opav{\f{\tilde A_{\cal A}}(\omega)}{[t,t+\tau]} &=& \frac{1}{\tau}\int\limits_t^{t+\tau} \f{A_{\cal A}}(t') e^{+i\omega t'} dt'\nn
&=& \frac{1}{\tau}\int\limits_t^{t+\tau} U^\dagger(t') A_{\cal A} U(t') e^{+i\omega t'} dt'\,,\nn
\eea
in the adiabatic limit.


\section{Spin-Boson Model}\label{Sspinboson}

We will make our method explicit at the example of the spin-boson model in the following. 
We will also give some numerical solutions to the master equations used:
The eigenvalues of the density matrices have been calculated with the LAPACK package \cite{lapack1999}.
The half-sided Fourier transforms (\ref{Eft_half}) and odd Fourier transforms (\ref{Eevenoddft}) were calculated numerically 
from the full Fourier transform by consecutive application of backward and forward integral 
transforms. This was implemented by an integration algorithm based on the discrete Fourier transform optimized for 
oscillating integrands \cite{press1994}.
For the integration of partial differential equations, a forth order Runge-Kutta method with an adaptive 
stepsize \cite{press1994} was used. Trace and hermiticity of the density matrix were always preserved within machine accuracy.
Likewise, whenever a Lindblad-type master equation was integrated, non-negativity of the smallest 
eigenvalue of the density matrix was preserved within numerical accuracy defined by the accuracy of the Fourier transform.


\subsection{Microscopic Derivation}\label{SSmicroscopic}

We consider a system Hamiltonian with discrete energy eigenvalues that can be described by a quadratic form of Pauli matrices for a system of $n$ spins
\bea\label{Espinboson}
\HS &=& \gamma \f{1} + \sum_{i=1}^n \left[\gamma^x_i \sigma^x_i + \gamma^y_i \sigma^y_i + \gamma^z_i \sigma^z_i\right]\nn
&&+ \sum_{i=1}^n \sum_{j=i+1}^n \sum_{\alpha,\beta=x,y,z} \gamma_{ij}^{\alpha\beta} \sigma_i^\alpha \sigma_j^\beta\,,
\eea
where $\vec\sigma_i$ denotes the Pauli matrices acting on spin $i$.
In the worst case, this Hamiltonian is defined by $1-3/2 n + 9/2 n^2$ real parameters.
Note that when considering explicit examples, we will not give the dimension of these parameters which implies that all
times have their inverse dimensions.
This system Hamiltonian is non-trivial in the sense that even in the case of time-independent parameters considered here,
the time evolution of the operators in the interaction picture cannot be solved analytically without using exponential resources.
The system is coupled to a bosonic bath
\bea\label{Ehbath_ex}
\HB = \sum_k \omega_k \left(b^\dagger_k b_k + \frac{1}{2}\right)
\eea
with the usual bosonic commutation relations.
The coupling between system and bath is realized by the quite general interaction
Hamiltonian
\bea\label{Ehint_ex}
\HI = \lambda \sum_{i=1}^n \sum_k \left[ \vec{n}_{ik} \cdot \vec{\sigma}_i \otimes b_k + \vec{n}_{ik}^* \cdot \vec{\sigma}_i \otimes b_k^\dagger\right]\,,
\eea
with $\lambda \ll 1$ where 
the frequency-dependence is contained in the complex-valued
coupling coefficients $\vec{n}_{ik}$.


\subsubsection{Bath Correlation Functions}

In order to obtain a rather simple form of the master equation, we will make the assumption of a collective coupling, 
where the frequency dependence of the coupling strengths factorizes with the different spin positions and spin directions, i.e.,
\bea\label{Ecollcoup}
\vec{n}_{ik} = \vec{n}_i h_k
\eea
for some function $h_k$. This implies that the distance between the spins is smaller than
the correlation length of the reservoir oscillators.
This approximation is not crucial for the further procedure but
simplifies the resulting system of equations considerably, since the
coupling to the reservoir can be described by just two effective spin operators.
In this case, the interaction Hamiltonian (\ref{Ehint_ex}) can be written as
\bea
\HI = \lambda \Sigma_R \otimes \left[B + B^\dagger\right] + \lambda \Sigma_I \otimes i\left[B - B^\dagger\right]
\eea
with the composed operators
\bea
\Sigma_R &=& \sum_{i=1}^n {\Re}\left( \vec{n}_i\right) \cdot \vec{\sigma}_i \,,\qquad \Sigma_I = \sum_{i=1}^n {\Im}\left( \vec{n}_i\right) \cdot \vec{\sigma}_i\,,\nn
B &=& \sum_k h_k b_k\,,
\eea
where we can identify from Eqn. (\ref{Eintham}) the hermitian operators
\bea
A_1 &=& \Sigma_R\,, \qquad A_2 = \Sigma_I\,,\nn
B_1 &=& B + B^\dagger\,, \qquad B_2 = i\left(B-B^\dagger\right)\,.
\eea
From 
the bath Hamiltonian (\ref{Ehbath_ex}) we obtain
$e^{i \HB \tau} b_k e^{-i \HB \tau} = e^{-i \omega_k \tau} b_k$ and the hermitian conjugate, respectively.
We will consider the limit of a large bath with a continuous spacing of oscillator frequencies. In 
this limit, the sums over $k$ can be approximated by an integral
\bea
\sum_k \abs{h_k}^2 f(\omega_k) \to \int\limits_0^\infty g(\omega) f(\omega) d\omega\,,
\eea
where $g(\omega)$ is defined by the distribution of bath oscillators (spectral density) as well as 
the frequency-dependent coupling strengths $h_k$.
With the bosonic expectation value for a thermalized reservoir 
\mbox{$\expval{N_k} = \traceB{b_k^\dagger b_k \RB^0} = \left[\exp\{\beta\omega_k\} - 1\right]^{-1}$}
at temperature $\beta=(k_{\rm B} T)^{-1}$
this can be used to determine the bath correlation functions as
\bea\label{Ecorr_func_spinboson}
C_{11}(\tau) 
&=& \frac{1}{2\pi} \int\limits_{-\infty}^{+\infty} \frac{2\pi g(\abs{\omega})}{\abs{1-e^{-\beta\omega}}} e^{-i\omega \tau} d\omega\,,\nn
C_{12}(\tau) 
&=& \frac{1}{2\pi} \int\limits_{-\infty}^{+\infty} \frac{2\pi (-i) g(\abs{\omega}){\rm sgn}(\omega)}{\abs{1-e^{-\beta\omega}}} e^{-i\omega \tau} d\omega\,,\nn
C_{21}(\tau) &=& -C_{12}(\tau)\,,\qquad
C_{22}(\tau) = C_{11}(\tau)\,,
\eea
where we can directly read off the Fourier transform matrix 
\bea\label{Espinbosonft}
\gamma(\omega) = \frac{2\pi g(\abs{\omega})}{\abs{1-e^{-\beta\omega}}}
\left(
\begin{array}{cc}
1 & -i\, {\rm sgn}(\omega)\\ 
+i\, {\rm sgn}(\omega) & 1
\end{array}
\right)\,.
\eea
Note that it is positive semidefinite (compare Appendix \ref{Asecular}) at every $\omega$ ensuring 
positive evolution of the Lindblad form master Eqn. (\ref{Elindblad_rwa1}).

In some examples given later-on, we will phenomenologically parameterize \cite{brandes,nesi} the density of states as
\bea\label{Edos}
g(\omega) = \left(\frac{\omega}{\omega_{\rm ph}}\right)^s e^{-\omega/\omega_{\rm ct}}\,,
\eea
where the exponent $s$ determines the behavior near small frequencies, $\omega_{\rm ph}$ is some physical frequency of the bath and
$\omega_{\rm ct}$ is a cutoff-frequency necessary for normalization.


\subsubsection{Exact Solution for Pure Dephasing}\label{SSpdexact}

The limit of pure dephasing of a single qubit is defined by considering $n=1$ with
\bea
\gamma=\frac{1}{2}\,, \qquad \gamma^z=-\frac{1}{2}\,,
\eea
and all other coefficients to vanish in Eqn. (\ref{Espinboson}) as well as the simple coupling
$\vec{n} = (0,0,1)$ in Eqn. (\ref{Ehint_ex}) -- in this case also the collective coupling 
assumption (\ref{Ecollcoup}) becomes exact.
In this case 
the time evolution operator in the interaction picture (\ref{Etimoint}) can be calculated 
exactly \cite{unruh1995a,lidar2001a,breuer2002} to all orders and one obtains that in the eigenbasis of the system Hamiltonian
the diagonal elements of the density matrix remain unchanged and the off-diagonal elements simply decay as
(cf. Eqn. (82) in \cite{lidar2001a} in the limit of a continuous bath spectrum)
\bea\label{Edecay_exact}
\f{\rho_{01}}(t) &=& e^{-\Gamma(t)} \f{\rho_{01}}(0)\,,\nn
\Gamma(t) &=& 8\lambda^2 \int\limits_0^\infty g(\omega) \frac{\sin^2(\omega t/2)}{\omega^2} {\coth}\left[\frac{\beta\omega}{2}\right] d\omega\,,
\eea
i.e., in the pure dephasing limit one does not obtain thermalization.
Likewise, since $e^{+i \HS t} A e^{-i \HS t} = A$ it is evident that BM and BMS approximations are equivalent 
for this case (compare also appendices \ref{Amarkov} and \ref{Asecular}).


\subsection{Non-Markovian Solutions}\label{SSexpdecay} 

For simple system and interaction Hamiltonians
\bea
\HS = \frac{1}{2} \left[\f{1} - \sigma^z\right]\,,\qquad
\HI = \lambda \sigma^{a} \otimes B_1\,,
\eea
where $a\in\{x,y,z\}$ and $B_1=B_1^\dagger$ is a bath operator, the Non-Markovian 
Born Eqn. (\ref{Eborn1}) is analytically solvable \cite{yu2000a,kleinekathoefer2004a,schroeder2007a}
in the special case of an exponentially decaying correlation function
\bea\label{Eexpdecay}
C_{11}(\tau) \equiv \traceB{\f{B_1}(\tau) B_1 \RB^0} = \frac{1}{2\tau_{\rm b}} e^{-\abs{\tau}/\tau_{\rm b}}\,.
\eea
By considering large $\tau_{\rm b}$ one can thus model reservoirs with a long-term memory, 
and the BM limit (where the BM-approximation becomes exact) is obtained by 
considering $\lim\limits_{\tau_{\rm b}\to 0} C_{11}(\tau)=\delta(\tau)$.
We obtain for the even and odd Fourier transforms in Eqn. (\ref{Eevenoddft})
\bea\label{Eevenoddft_01}
\gamma_{11}(\omega) = \frac{1}{1+\left(\omega \tau_{\rm b}\right)^2}\,,\qquad
\sigma_{11}(\omega) = \frac{i \omega \tau_{\rm b}}{1+\left(\omega \tau_{\rm b}\right)^2}\,,
\eea
where it becomes visible (for later comparison with the spin-boson model) from 
Eqn. (\ref{Ecorr_func_spinboson}) that this 
case can in principle be reproduced 
by a bosonic reservoir in the large temperature limit with a Drude-like slowly decaying 
(but temperature-dependent) spectral coupling density
\bea\label{Especdens_special}
g(\omega) = \frac{1}{2\pi} \frac{\beta \omega}{1+\left(\omega\tau_{\rm b}\right)^2}\,.
\eea
Note however that by assuming a sum of many exponentials and allowing for phases, 
the method can in principle be generalized 
also to the low-temperature limit  \cite{kleinekathoefer2004a,burkard08030564}.
Inserting the operator definitions in the Born Eqn. (\ref{Eborn1}) we obtain
\bea\label{Eexample_a}
\eta_a(t) &\equiv& \int\limits_0^t e^{+\frac{i}{2}(t-t') \sigma^z} \sigma^a \RS(t') e^{-\frac{i}{2}(t-t') \sigma^z} \frac{e^{-(t-t')/\tau_{\rm b}}}{2\tau_{\rm b}} dt'\,,\nn
\dot{\RS} &=& \frac{i}{2} \left[ \sigma^z, \RS(t)\right]
+ \lambda^2 \left[\eta_a(t) - \eta_a^\dagger(t), \sigma^a\right]\,,
\eea
where it becomes evident that by taking the time derivative of the operator $\eta^a$ one simply obtains a coupled set of first order
differential equations  
\bea\label{Ecoupled}
\dot{\RS} &=& \frac{i}{2} \left[ \sigma^z, \RS(t)\right] + \lambda^2 \left[\bar \eta_a(t), \sigma^a\right]\,,\nn
\dot{\bar\eta}_a &=& \frac{i}{2} \left[\sigma^z, \bar\eta_a(t)\right] - \frac{1}{\tau_{\rm b}} \bar\eta_a(t) + \frac{1}{2\tau_{\rm b}} \left[\sigma^a, \RS(t)\right]\,.
\eea
for the operators $\bar\eta_a(t)=\eta_a(t)-\eta_a^\dagger(t)$ and $\RS(t)$.
Evidently, trace and hermiticity of $\RS$ and anti-hermiticity of $\bar\eta_a$ are always preserved. Due to the initial 
condition $\bar\eta_a(0)=\f{0}$, the trace of $\bar\eta_a$ will always vanish.
Therefore, it suffices to parameterize $\RS$ and $\bar \eta$ by just six real variables
\bea\label{Eparametrization}
\RS &\equiv& \left(\begin{array}{cc}
\rho_{00} & \rho_x + i \rho_y\\
\rho_x - i \rho_y & 1 - \rho_{00}
\end{array}
\right)\,,\nn
\bar\eta &\equiv& \left(\begin{array}{cc}
i \eta_{00} & \eta_x + i \eta_y\\
-\eta_x + i \eta_y & - i \eta_{00}
\end{array}
\right)\,.
\eea
In the limit $\tau_{\rm b}\to 0$ we simply obtain
\bea\label{Eexpmarkov}
\dot{\RS} = \frac{i}{2}\left[ \sigma^z, \RS(t)\right] + \lambda^2 \left[\sigma^a \RS(t) \sigma^a - \RS(t)\right]
\eea
for the system density matrix.
In the BM approximation (with finite $\tau_{\rm b}$) we obtain along the lines of Appendix \ref{Amarkov}
for $a=z$ (pure dephasing)
$\eta_z^{\rm BM}(t) = \frac{1}{2} \sigma^z \RS(t)$ and thereby
\bea\label{Eexpmarkovzapp}
\dot{\RS}^{\rm BM} =  \frac{i}{2}\left[ \sigma^z, \RS(t)\right] + \lambda^2 \left[\sigma^z \RS(t) \sigma^z - \RS(t)\right]
\eea
which coincides with (\ref{Eexpmarkov}), i.e., the dependence on $\tau_{\rm b}$ vanishes.
In contrast, for the more interesting dissipative case ($a=x$) we obtain
\mbox{$\eta_x^{\rm BM}(t) = \frac{1}{2}\frac{1}{1+\tau_{\rm b}^2} \sigma^x \RS(t) - \frac{1}{2} \frac{\tau_{\rm b}}{1+\tau_{\rm b}^2} \sigma^y \RS(t)$},
which yields
\bea\label{Eexpmarkovxapp}
\dot{\RS}^{\rm BM} &=&  \frac{i}{2}\left[ \sigma^z, \RS(t)\right] + \frac{\lambda^2}{2} \frac{1}{1+\tau_{\rm b}^2} \left[\left[\sigma^x, \RS(t)\right], \sigma^x\right]\nn
&&- \frac{\lambda^2}{2} \frac{\tau_{\rm b}}{1+\tau_{\rm b}^2} \left[\left[\sigma^y, \RS(t)\right], \sigma^x\right]\,.
\eea

\subsubsection{Pure Dephasing}

In the pure dephasing case one has $a=z$. 
Inserting the matrix elements in (\ref{Ecoupled}) one finds that
$\RS^{00}$ and $\RS^{11}$ are time-independent (as is also known from the full solution) 
and that the time evolution of the off-diagonal matrix element is governed by
\bea
\frac{d}{dt}
\left(\begin{array}{c}
\RS^{01}(t)\\
\bar\eta^{01}(t)
\end{array}\right)
= \left(\begin{array}{cc}
i & -2\lambda^2\nn
+1/\tau_{\rm b} & i - 1/\tau_{\rm b}
\end{array}\right)
\left(\begin{array}{c}
\RS^{01}(t)\\
\bar\eta^{01}(t)
\end{array}\right)\,.
\eea
With the initial condition $\bar\eta^{01}(0)=0$ one finds
\bea\label{Edecay_01}
\RS^{01}(t) &=& \RS^{01}(0) e^{it} e^{-t/(2\tau_{\rm b})} \Big[
\cosh\left(\sqrt{1-8\lambda^2\tau_{\rm b}}\frac{t}{2\tau_{\rm b}}\right)\nn
&&+ \frac{1}{\sqrt{1-8\lambda^2\tau_{\rm b}}}\sinh\left(\sqrt{1-8\lambda^2\tau_{\rm b}}\frac{t}{2\tau_{\rm b}}\right)\Big]\,,
\eea
which reproduces the decay of the off-diagonal elements (the factor $e^{it}$ is a consequence of the Schr\"odinger
picture). In the high-temperature limit and for the corresponding density of states (\ref{Especdens_special}) leading to exponentially decaying 
correlation functions (\ref{Eexpdecay}), the decay rate of the exact solution (\ref{Edecay_exact}) becomes
\bea
\Gamma(t) &=& \frac{8\lambda^2}{\pi} \int\limits_0^\infty \frac{\sin^2(\omega t/2)}{\omega^2\left[1+(\omega\tau_{\rm b})^2\right]} d\omega\nn
&=& 4\lambda^2\tau_{\rm b} \left[ \cosh^2\left(\frac{t}{2\tau_{\rm b}}\right) - 1\right.\nn
&&\left. - \sinh\left(\frac{t}{2\tau_{\rm b}}\right) \cosh\left(\frac{t}{2\tau_{\rm b}}\right)
+ \frac{t}{2\tau_{\rm b}}\right]\,,
\eea
which reduces in the limit $\tau_{\rm b}\to 0$ to $\Gamma(t)\approx 2\lambda^2 t$. Similarly, we find that in this limit, 
Eqn. (\ref{Edecay_01}) reduces to $\RS^{01}(t) = \RS^{01}(0) e^{it} e^{-2\lambda^2 t}$. 
This can be understood as the limit $\tau_{\rm b}\to 0$ also corresponds to an infinitely fast relaxation 
of the reservoir, where also the Born approximation becomes exact.
Likewise, it is straightforward to see that $\abs{\RS^{01}(t)}-e^{-\Gamma(t)}\RS^{01}(0) = \ord\{\lambda^4\}$.
We will therefore not further discuss the pure dephasing case with exponentially decaying correlation functions
and compare with the exact solution (\ref{Edecay_exact}) instead later-on.

\subsubsection{Dissipative Coupling}

Another important case is reproduced by choosing the dissipation coupling $a=x$ in Eqn. (\ref{Eexample_a}). 
Inserting the ansatz (\ref{Eparametrization}) into Eqn. (\ref{Ecoupled}) one finds two $3 \times 3$ systems
\bea\label{Esolborn}
\left(\begin{array}{c}
\dot\rho_{00}\\
\dot\eta_x\\
\dot\eta_y
\end{array}\right) &=& 
\left(\begin{array}{ccc}
0 & 2\lambda^2 & 0\\
-\tau_{\rm b}^{-1} & -\tau_{\rm b}^{-1} & -1\\
0 & 1 & -\tau_{\rm b}^{-1}
\end{array}\right)
\left(\begin{array}{c}
\rho_{00}\\
\eta_x\\
\eta_y
\end{array}\right)\nn
&&+\frac{1}{2\tau_{\rm b}} 
\left(\begin{array}{c}
0\\
1\\
0
\end{array}\right)\,,\nn
\left(\begin{array}{c}
\dot\rho_x\\
\dot\rho_y\\
\dot\eta_{00}
\end{array}\right) &=& 
\left(\begin{array}{ccc}
0 & -1 & 0\\
1 & 0 & 2\lambda^2\\
0 & -\tau_{\rm b}^{-1} & -\tau_{\rm b}^{-1}
\end{array}\right)
\left(\begin{array}{c}
\rho_x\\
\rho_y\\
\eta_{00}
\end{array}\right)\,,
\eea
which have an analytic solution that is too lengthy to be 
reproduced here. At first glance, these systems seem to be completely independent but note that
the condition of initial validity of the density matrix relates their initial conditions.
The steady-state solution for the density matrix corresponds to the
thermalized Gibbs state (\ref{Egibbs}) for high temperatures ($\beta\to 0$).


\subsection{Single-Qubit Coarse Graining}

\subsubsection{Pure Dephasing}

\begin{figure}
\includegraphics[height=6cm]{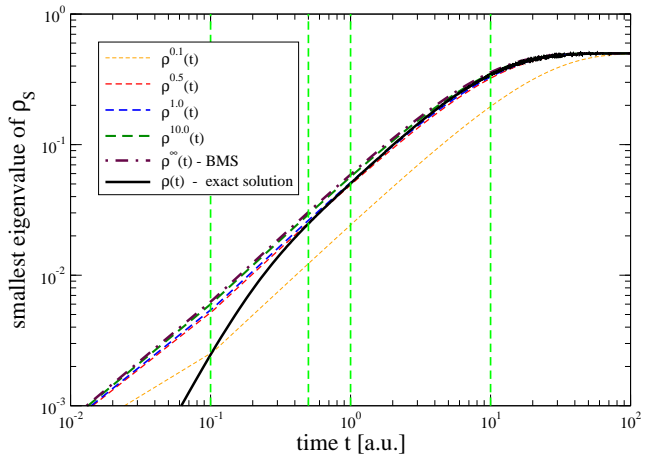}
\caption{\label{Fsqbit_dephasing}[Color Online]
Eigenvalue evolution of the system density matrix (only one eigenvalue shown since trace-conservation implies a symmetric evolution) 
initialized in the pure state $\bra{i}\RS^0\ket{j}=1/2$ for single-qubit dephasing.
As predicted by Eqn. (\ref{Eintersect_dephasing}), the solutions of the coarse-graining master equations (thin dashed lines)
intersect the exact solution (solid line) at $t=\tau$ (vertical dashed lines). 
The secular approximation (thick dashed line) corresponds to $\tau\to\infty$ and does not correctly cover
the short time behavior of the exact solution. Parameters
were chosen as follows: $\beta = 1$, $\omega_{\rm ph}=1$, $\omega_{\rm ct}=5$, $s=1$, $\lambda=0.1$.
}
\end{figure}

In order to determine the master equations (\ref{Emaster_01}) for the pure dephasing limit
discussed in subsection \ref{SSpdexact}, we use 
\bea
A_1 &=& \sigma^z \,,\qquad B_1 = \sum_k \left[h_k b_k + h_k^* b_k^\dagger\right]
\eea
to obtain that the Lamb shift Hamiltonian in Eqn. (\ref{Eliouvillian}) is proportional to the identity matrix
and thus has no effect.
From the dissipative part however we obtain a non-vanishing contribution from Eqn. (\ref{Eliouvillian}),
such that the master Eqn. (\ref{Emaster_replacement}) in the interaction picture is given by
\bea
\f{\dot\RS^\tau} &=& \bar \Gamma(\tau) \Big[ \ket{0} \bra{0} \f{\RS^\tau}(t) \ket{0} \bra{0} + \ket{1} \bra{1}\f{\RS^\tau}(t) \ket{1} \bra{1}\nn
&&-\ket{0} \bra{0} \f{\RS^\tau}(t) \ket{1} \bra{1} - \ket{1} \bra{1}\f{\RS^\tau}(t) \ket{0} \bra{0}\Big]\nn
&&-\bar \Gamma(\tau) \f{\RS^\tau}(t)\,,\nn
\bar\Gamma(\tau) &=& \frac{\lambda^2\tau}{2\pi} \int\limits_{-\infty}^{+\infty} \gamma_{11}(\omega) {\rm sinc}^2\left[\frac{\omega \tau}{2}\right]  d\omega\,.
\eea
With assuming exponentially decaying correlation functions (\ref{Eexpdecay}) we can use Eqn. (\ref{Eevenoddft_01}) to recover the 
BM approximation (\ref{Eexpmarkovzapp}), but above equation for pure dephasing of course holds also for more general correlation functions.
It is straightforward to show that the off-diagonal elements of the density matrix will decay as
\mbox{$\bra{0} \f{\RS^\tau}(t) \ket{1} = e^{-\bar \Gamma(\tau) t}  \bra{0}\f{\RS^\tau}(0)\ket{1}$}.
A closer inspection of the decay rate yields
\bea\label{Eintersect_dephasing}
\bar \Gamma(\tau) t &=& 8 \lambda^2 \frac{t}{\tau} \int\limits_{-\infty}^{+\infty} \frac{\sin^2(\omega \tau/2)}{\omega^2} \frac{g(\abs{\omega})}{\abs{1-e^{-\beta \omega}}} d\omega\nn
&=& 8 \lambda^2 \frac{t}{\tau} \int\limits_{0}^{+\infty} \frac{\sin^2(\omega \tau/2)}{\omega^2} g(\omega) {\rm coth}(\beta \omega/2) d\omega\nn
&=&  \frac{t}{\tau} \Gamma(\tau)\,,
\eea
i.e., we reproduce the result of \cite{lidar2001a} that for pure dephasing of a single qubit, the solution $\RS^t(t)$ yields the exact 
solution $\RS(t)$, see also figure \ref{Fsqbit_dephasing}.
Naturally, this is also equivalent with the Born Equation up to $\ord\{\lambda^2\}$.
Note also that for large coarse graining times, figure \ref{Fsqbit_dephasing} shows that the coarse-graining master equations approach the secular approximation
master equation -- which can also be seen as a numerical confirmation of Eqn. (\ref{Esincidentity}).


\subsubsection{Dissipative Coupling}

With considering $A_1 = \sigma^x$, an exponentially decaying correlation function (\ref{Eexpdecay}), and
$\HS = \frac{1}{2}\left[\f{1} - \sigma^z\right]$ one obtains a more complicated master equation for $\RS^\tau(t)$
in the Schr\"odinger picture
\bea\label{Esolcoarsegraining}
\dot{\RS^\tau}(t) &=& 
\left[\frac{i}{2}\sigma^z -i \sigma_{00}(\tau) \ket{0}\bra{0} -i \sigma_{11}(\tau) \ket{1}\bra{1}, \RS^\tau\right]\nn
&&+\gamma_{01, 01}(\tau) \left[ \ket{0} \bra{1} \RS^\tau \ket{1}\bra{0} - \frac{1}{2} \left\{\ket{1}\bra{1}, \RS^\tau\right\}\right]\nn
&&+\gamma_{10, 10}(\tau) \left[ \ket{1} \bra{0} \RS^\tau \ket{0}\bra{1} - \frac{1}{2} \left\{\ket{0}\bra{0}, \RS^\tau\right\}\right]\nn
&&+\gamma_{01, 10}(\tau) \ket{0}\bra{1} \RS^\tau \ket{0}\bra{1}\nn
&&+\gamma_{10, 01}(\tau) \ket{1}\bra{0}\RS^\tau\ket{1}\bra{0}\,.
\eea
From the non-vanishing matrix elements in Eqn. (\ref{Emaster_replacement}) one can deduce that (unlike in the pure dephasing case) the
Lamb-shift term will contribute, since (although diagonal) it is not 
proportional to the identity matrix anymore.
Likewise, we also observe here a decoupled evolution of diagonal 
\bea
{\RS^\tau}_{00}(t) &=& \frac{\gamma_{01,01}}{\gamma_{01,01}+\gamma_{10,10}}\left[1-e^{-\left(\gamma_{01,01}+\gamma_{10,10}\right)t}\right]\nn
&&+ e^{-\left(\gamma_{01,01}+\gamma_{10,10}\right)t} {\RS^\tau}_{00}(0)
\eea
and off-diagonal matrix elements
\bea
{\dot{\RS}^{\tau}}_{01} &=& \left[i - \Gamma(\tau)\right] {\RS^\tau}_{01} + \gamma_{01,10}(\tau) \left({\RS^\tau}_{01}\right)^*\,, 
\eea
with $\Gamma(\tau)=\frac{1}{2}\left(\gamma_{01,01}+\gamma_{10,10}\right) - i\left(\sigma_{11}-\sigma_{00}\right)$ (we have omitted the $\tau$-dependence).
Considering even and odd Fourier transforms of the form (\ref{Eevenoddft_01}) corresponding to exponentially decaying correlation functions, 
we can now compare the solution (\ref{Esolborn}) of the Born equation
with the solutions (\ref{Esolcoarsegraining}) of the coarse-graining approach.

\begin{figure}
\includegraphics[height=6cm]{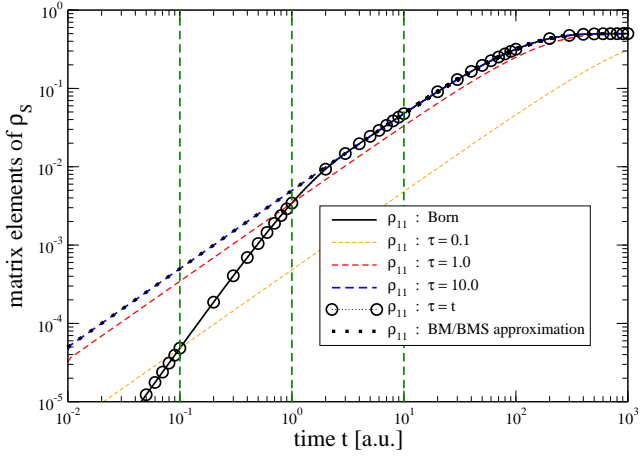}
\caption{\label{Fdecay_sqbitdissip_rho11}[Color Online]
Evolution of the $\rho_{11}$-matrix element of the density matrix for dissipative coupling.
The solid line represents the solution of the Born Eqn. (\ref{Esolborn}) and
the thin dashed lines correspond to solutions of (\ref{Esolcoarsegraining}) for different coarse graining
times $\tau$. Up to second order in $\lambda$, a perfect numerical agreement is found for the matrix 
elements of $\rho^t(t)$ (symbols) and the Born solution (solid line). 
Parameters were chosen as follows: $\lambda=0.1$, $\tau_{\rm b}=1$.
}
\end{figure}

When one initializes the density matrix as \mbox{$\RS(0)=\ket{0}\bra{0}$}
one finds that for infinite coarse-graining times the diagonal terms will equilibrate 
to value 1/2 (which corresponds to thermalization at infinite temperatures), 
see figure \ref{Fdecay_sqbitdissip_rho11}.
In this case, the off-diagonal terms will evidently vanish throughout. 
Note that for the small coupling chosen ($\lambda=0.1$), the solution
of the Born equation $\RS(t)$ is approximated by the adaptive coarse-graining solutions $\RS^t(t)$ with extraordinary accuracy.

\begin{figure}
\includegraphics[height=6cm]{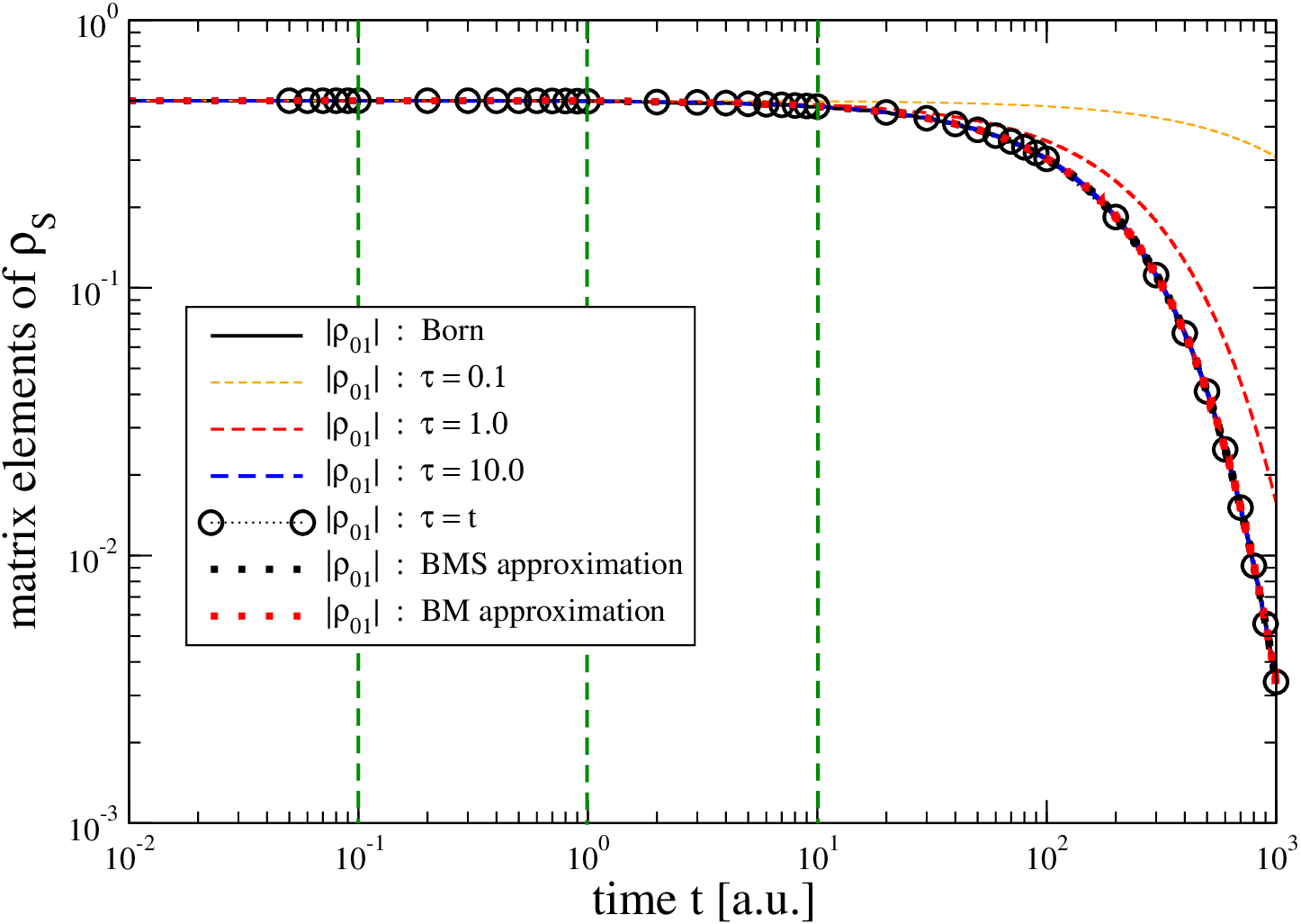}
\caption{\label{Fdecay_sqbitdissip_rho01}[Color Online]
Evolution of the $\abs{\rho_{01}}$ matrix element of the density matrix for pure dissipation.
The solid line represents the solution of the Born Eqn. (\ref{Esolborn}) and
the thin dashed lines correspond to solutions of (\ref{Esolcoarsegraining}) for different coarse graining
times $\tau$. Up to second order in $\lambda$, a perfect agreement is found for the matrix elements 
of $\rho^t(t)$ (symbols) and the Born solution (solid line).
The difference between Born solution and  BM as well as BMS approximations is too small to be visible.
Parameters were chosen as follows: $\lambda=0.1$, $\tau_{\rm b}=1$.
}
\end{figure}

In contrast, when initializing the density matrix as
$\RS(0) = \frac{1}{2} \left[\ket{0}\bra{0} + \ket{0}\bra{1}+\ket{1}\bra{0}+\ket{1}\bra{1}\right]$
one observes that the diagonal entries will remain unchanged and the off-diagonals will decay.
The actual behavior of the decay of the off-diagonal elements is depicted in figure \ref{Fdecay_sqbitdissip_rho01}.
Again we observe a strikingly good agreement between the Born solution $\RS(t)$ and the adaptive coarse-graining
solutions $\RS^t(t)$.

It is also instructive to compare for the adaptive coarse-graining approach the limit of infinite coarse-graining times 
(BMS approximation)
\bea
\dot{\RS} &=& \left(\frac{i}{2} + \frac{i}{2} \frac{\lambda^2\tau_{\rm b}}{1+\tau_{\rm b}^2}\right) \left[\sigma^z, \RS\right]\nn
&&+\frac{\lambda^2}{1+\tau_{\rm b}^2}\left[\ket{0}\bra{1} \RS\ket{1}\bra{0} + \ket{1}\bra{0} \RS\ket{0}\bra{1} - \RS\right]\nn
\eea
with the BM approximation master equation (\ref{Eexpmarkovxapp}), where one can see that only the equations for the diagonals match.
Likewise, in the limit $\tau_{\rm b}\to 0$ one obtains
\bea
\dot{\RS^\tau} &=& \frac{i}{2} \left[\sigma^z, \RS^\tau\right]\nn
&&+\lambda^2\left[\ket{0}\bra{1}\RS^\tau\ket{1}\bra{0} + \ket{1}\bra{0}\RS^\tau\ket{0}\bra{1} - \RS^\tau\right]\nn
&&+\lambda^2 e^{-i\tau} \sinc{\tau} \ket{0}\bra{1}\RS^\tau\ket{0}\bra{1}\nn
&&+\lambda^2 e^{+i\tau} \sinc{\tau} \ket{1}\bra{0}\RS^\tau\ket{1}\bra{0}\,.
\eea
Here, one finds that the BMS approximation ($\tau\to\infty$) yields a different master equation than the BM 
limit (\ref{Eexpmarkov}) and also the BM approximation (\ref{Eexpmarkovxapp}) 
However, also within the coarse-graining approach the limit $\tau_{\rm b}\to 0$ (with finite $\tau$) leads to the same 
steady state as the BMS approximation $\tau \to \infty$ and the BM limit (\ref{Eexpmarkov}) -- only 
the relaxation rates may differ.
Note however that the BM approximation (\ref{Eexpmarkovxapp}) may even lead to instable behavior of the off-diagonal 
matrix elements for large $\lambda$ (compare also section \ref{SSpositive}), whereas the coarse-graining approach always generates Lindblad forms.


\subsubsection{General Coupling}

\begin{figure}[ht]
\includegraphics[height=6cm]{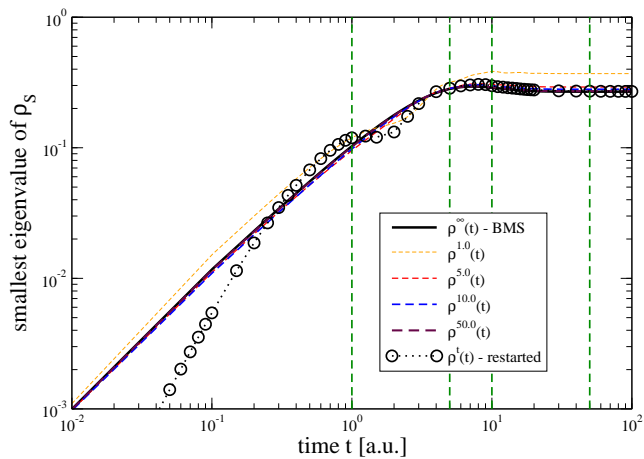}
\caption{\label{Fgraining}[Color Online]
Evolution of the eigenvalues of the density matrix for a single qubit initialized in the pure state $\bra{i}\RS^0\ket{j}=1/2$. 
For long times the BMS solution (thick dashed line) and the adaptive coarse-graining solution (thick solid line) predict
thermalization in the thermal equilibrium state (\ref{Egibbs}), whereas a different equilibrium state is assumed for short coarse-graining 
times (thin dashed lines).
Parameters were chosen as follows: $\beta = 1$, $\omega_{\rm ph}=1$, $\omega_{\rm ct}=5$, $s=1$, $\lambda=0.1$.}
\end{figure}

For a more complex coupling of the spin to the reservoir 
\bea\label{Ecoup_01}
\HS = \frac{1}{2}\left[\f{1} - \sigma^z\right]\,,\qquad \vec{n}_1 = \frac{1}{\sqrt{2}}(1+i, 1+i, 1+i)\,,
\eea
and also more realistic spectral coupling densities of type (\ref{Edos}) 
we find thermalization as predicted by the BMS approximation for long times. 
For short times however, the solution $\RS^t(t)$ may strongly differ from the BMS solution, see figure \ref{Fgraining}.
For example, if coarse-graining times are chosen too small, a non-thermalized steady state is obtained.
For large coarse-graining times, the thermalized steady state is reached, but then in the short time regime, large
differences between fixed graining and the adaptive graining solution are found.
Again, we numerically confirm Eqn. (\ref{Esincidentity}), since for large coarse graining times the secular approximation
is reproduced.


\subsection{Staying Reasonable}\label{SSpositive}

Frequently, the BM approximation (\ref{Emaster_mknrwa}) is used although it does not generally guarantee for positive evolution.
This is certainly tolerable if the solution obtained approximates the exact solution well (and thus violates positivity only slightly).
In addition, if the interaction Hamiltonian used in such models implicitly corresponds to a secular approximation \cite{childs2001,schuetzhold2005a}, one
will obtain a Lindblad form master equation and positivity as well as stability of the density matrix will be granted throughout. 
In general however, this will not be the case. 
Here we will analytically consider a single qubit
$\HS=\frac{1}{2}\left[\f{1}-\sigma^z\right]$ with the simple coupling
$\HI=\sigma^x \otimes B$.
Denoting with $\Gamma(\omega)$ the half-sided Fourier-transform (\ref{Eft_half}) of the reservoir correlation function,
one can write Eqn. (\ref{Emaster_mknrwa}) in the form $\dot{\RS}=L\RS(t)$ and calculate the eigenvalues of the Liouvillian.

If one is only interested in the subspace of diagonal density matrices one finds the 
two corresponding eigenvalues $\sigma_0=0$ and $\sigma_1=-2\lambda^2 A$ with 
\mbox{$A=\Re \left\{\Gamma(+1)+\Gamma(-1)\right\}$}.
For physically motivated bath correlation functions $C_{11}$ in Eqn. (\ref{Ecorr_func_spinboson}) one obtains 
that $\sigma_1 < 0$, such that the evolution in this subspace may not lead to unstable behavior -- although positivity may be violated.

\begin{figure}[ht]
\includegraphics[height=6cm]{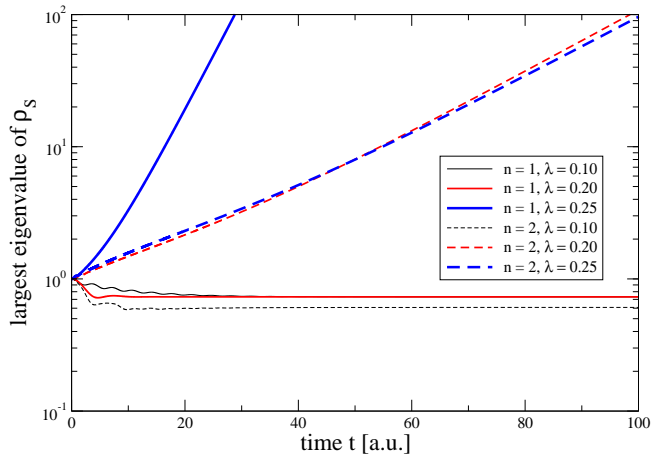}
\caption{\label{Fmarkov_failure2}[Color Online]
Largest eigenvalue of the density matrix for one (solid) and two (dashed) qubits initialized with $\bra{i}\RS^0\ket{j}=1/2^n$.
calculated from Eqn. (\ref{Emaster_mknrwa}).
For large $\lambda$, not only positivity is violated (negative eigenvalues not shown -- trace conservation), 
but the solution even becomes unstable.
The parameters in (\ref{Ecorr_func_spinboson}) were chosen as $\beta = 1$, $\omega_{\rm ph}=1$, $\omega_{\rm ct}=5$, $s=2$, 
such that the transition from stable to unstable behavior occurs at $\lambda_{\rm crit}\approx 0.231294$ for $n=1$.
}
\end{figure}

In contrast, in the off-diagonal subspace one obtains the two eigenvalues
$\lambda_{2/3}=-2\lambda^2 A \pm \sqrt{-1 + 2 B \lambda^2 + A^2 \lambda^4}$
with $A$ defined as above and $B = \Im\left\{\Gamma(-1)-\Gamma(+1)\right\}$.
Given $B>0$ (which can be achieved with correlation functions of form (\ref{Ecorr_func_spinboson})), 
one of these will pick a positive real part as soon as $\lambda^2 > (2 B)^{-1}$, 
which corresponds to an unstable evolution, see also figure \ref{Fmarkov_failure2}.
Numerically, we observe the same for $n=2$ mutually uncoupled qubits (with similar system Hamiltonians and couplings).
In this case, the expectation value of operators that are not diagonal in the system Hamiltonian will 
not yield any meaningful results and the Born-Markov approximation is questionable.

\subsection{Thermalization}

In the pure dephasing limit, one observes a rapid (i.e., exponential) decay of the off-diagonal elements of the density matrix. Also, if there 
are no degeneracies in the spectrum, the rotating wave approximation (\ref{Elindblad_rwa1}) predicts a decoupled evolution of the diagonal elements
of the density matrix in the eigenbasis of $\HS$, i.e., from the corresponding evolution equation $\dot{\rho}_{ii}(t)=\sum_j A_{ij} \rho_{jj}(t)$ one might expect 
an exponential decay into the eigenvector of the matrix $A$ which has eigenvalue 0 (steady state). 
The corresponding subspace can still be degenerate (many stationary states) or there may exist exponentially many
other eigenvectors with very small eigenvalues, such that thermalization does not necessarily happen very fast.
Here we will consider some specific realizations of
the spin-boson model (\ref{Espinboson}) and solve Eqn. (\ref{Elindblad_rwa1}) numerically.

\begin{figure}[ht]
\includegraphics[height=6cm]{statstate.eps}\\
\begin{tabular}{ccccc}
\includegraphics[height=1.5cm]{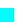} &
\includegraphics[height=1.5cm]{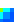} &
\includegraphics[height=1.5cm]{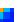} &
\includegraphics[height=1.5cm]{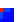} &
\includegraphics[height=1.5cm]{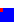}\\
\includegraphics[height=1.5cm]{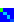} &
\includegraphics[height=1.5cm]{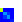} &
\includegraphics[height=1.5cm]{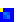} &
\includegraphics[height=1.5cm]{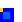} &
\includegraphics[height=1.5cm]{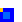}\\
\includegraphics[height=1.5cm]{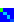} &
\includegraphics[height=1.5cm]{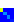} &
\includegraphics[height=1.5cm]{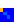} &
\includegraphics[height=1.5cm]{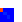} &
\includegraphics[height=1.5cm]{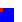}
\end{tabular}
\caption{\label{Fstatstate}[Color Online]
{\bf Top:}
Eigenvalue evolution for a 2-qubit system according to the secular approximation master Eqn. (\ref{Elindblad_rwa1}) with a two-fold degeneracy 
for different initial states (solid and dashed lines) and with a near two-fold degeneracy (dotted lines).
Vertical dashed lines indicate the times at which the bottom row snapshots have been taken.\\
{\bf Bottom rows:}
Absolute values of the matrix elements of the corresponding $4\times 4$ density matrix in the ordered energy eigenbasis (such that the top left corner corresponds to
the ground state). Color coding ranges from blue (value 0) to red (value 1). 
In the upper row, the system is initialized in a non-thermal density matrix $\bra{i}\RS^0\ket{j}=1/4$ and relaxes into a the thermalized state (\ref{Egibbs}) with $\beta_{\rm sys}=\beta$,
i.e., for the low reservoir temperature chosen ($\beta=10$) essentially into the ground state.
In contrast, in the middle row the initial state was a thermalized one with initial system temperature $\beta^0_{\rm sys}=1$, and the system does not relax into a 
thermal equilibrium state.
In the lowest row, the degeneracy was broken by choosing $\gamma^z_1 = -1/2+0.01$ and $\gamma^z_2 = -1/2-0.01$.
\\
Other parameters were chosen as
$\beta = 10$, $\omega_{\rm ph}=1$, $\omega_{\rm ct}=5$, $s=1$, $\lambda=0.1$.
}
\end{figure}

For example, if the system Hamiltonian has degeneracies and the 
coupling to the reservoir does not lift these, the system may relax into states that are not even diagonal in the system Hamiltonian basis. 
In these cases, the initial density matrix may determine which steady state is actually reached, see figure \ref{Fstatstate}.
As a first example we consider
\bea\label{Ethermalization}
\gamma=1\,, \qquad \gamma^z_1=\gamma^z_2 = -\frac{1}{2}\,,\nn
\vec{n}_1 =\vec{n}_2 = \frac{1}{\sqrt{2}}(1+i, 1+i, 1+i)\,,
\eea
and all other coefficients of Eqn. (\ref{Espinboson}) vanishing, such that there exists a two-fold degeneracy in the spectrum of $\HS$.
In this subsection, we will consider spectral densities of type (\ref{Edos}).
For the low reservoir temperatures assumed in figure \ref{Fstatstate}, thermalization corresponds to relaxation into the ground state and one can see that it may depend
on the initial state of the density matrix and possibly lifted degeneracies whether thermalization takes place. In reality, the degeneracy might always be lifted 
by some imperfect Hamiltonian implementations. In addition, the coupling to the reservoir may be more complex than in (\ref{Ethermalization}) thus making the thermalized system
state (\ref{Egibbs}) the only stationary state of the rotating wave master Eqn. (\ref{Elindblad_rwa1}), compare also \cite{breuer2002}. 
However, it is to be expected that thermalization in this case will
take rather long times, especially for large and complicated systems (see next subsection).


\subsection{Solving Problems by Cooling}

It is known that Hamiltonians of the form (\ref{Espinboson}) can be used to encode solutions to computationally hard problems in their ground state.
This is for example exploited in adiabatic quantum computation \cite{farhi2001a}. For a system made of five spins, we will compare an enlarged version of
the previous example (\ref{Ethermalization})
\bea\label{Eham_00}
\gamma=\frac{5}{2}\,, \qquad \gamma^z_{i=1..5} = -\frac{1}{2}
\eea
with the ground state encoding of {\em Exact Cover 3} -- a specific NP-complete problem. 

The Exact Cover 3 problem can be introduced as follows \cite{farhi2001a}:
Given a set of $m$ constraints where each constraint contains the positions of three bits $C_m=(p^1_m, p^2_m, p^3_m)$ (with evidently $1 \le p^i_m \le n$), one is looking for 
the $n$-bit bit-string $b_1, \ldots, b_n$ (with $b_i \in \{0,1\}$) that fulfills for each constraint $b_{p^1_m}+b_{p^2_m}+b_{p^3_m}=1$ (where ''+'' denotes the integer sum).
One way to encode the solution to this problem into the ground state of a Hamiltonian of the type (\ref{Espinboson}) is given by \cite{banuls2006a,schuetzhold2006a}
\bea\label{Eham_01}
\gamma=m\,,\qquad\gamma^z_i = - \frac{n_i}{2}\,,\qquad \gamma^{zz}_{ij}=+\frac{n_{ij}}{2}
\eea
and all other coefficients vanishing -- the differing pre-factor of $\gamma^{zz}_{ij}$ in 
comparison to \cite{schuetzhold2006a} results from the absence of double-counting in $ij$ in Eqn. (\ref{Espinboson}).
In above equation, $m$ denotes the total number of clauses, $n_i$ the number of clauses involving bit $i$, and 
$n_{ij}$ the number of clauses involving both bits $i$ and $j$.
Specifically, we will consider the 4 clauses
$C_1 = (2,3,4)$, $C_2 = (1,2,5)$, $C_3 = (1,4,5)$, and $C_4 = (3,4,5)$.
This implies the non-vanishing coefficients 
\mbox{$\gamma=4$}, \mbox{$\gamma^z_1 = \gamma^z_2 = \gamma^z_3 = -1$}, \mbox{$\gamma^z_4 = \gamma^z_5 = -3/2$},
\mbox{$\gamma^{zz}_{12}=\gamma^{zz}_{14}=\gamma^{zz}_{23}=\gamma^{zz}_{24}=\gamma^{zz}_{25}=\gamma^{zz}_{35}=1/2$}, and
\mbox{$\gamma^{zz}_{15}=\gamma^{zz}_{34}=\gamma^{zz}_{45}=1$}.

\begin{figure}[ht]
\includegraphics[height=6cm]{relaxation_ec3.eps}\\
\begin{tabular}{cc}
\includegraphics[height=3cm]{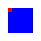} &
\includegraphics[height=3cm]{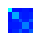}
\end{tabular}
\caption{\label{Frelaxation_ec3}[Color Online]
{\bf Top:}
Eigenvalue evolution for a 5-qubit system according to the secular approximation master Eqn. (\ref{Elindblad_rwa1}) for different system Hamiltonians
given by Eqn. (\ref{Eham_00}) (solid lines) and equation (\ref{Eham_01}) with an Exact Cover 3-problem (dashed lines), respectively. In both cases, the density
matrix has been initialized as $\bra{i}\RS^0\ket{j}=1/32$. Whereas for the 
uncoupled spin system the system rapidly relaxes into the ground state, the final state for the exact cover problem is very different.
{\bf Bottom:}
Absolute values of the matrix elements of the $7\times 7$ top left sub-matrix of the full $32 \times 32$ density matrix in the ordered energy eigenbasis.
On the left (corresponding to the solid line on top), the system relaxes to the ground state (red spot), whereas for the exact cover problem 
(right, dashed line on top), the system state is not thermalized.\\
Other parameters were chosen as
$\beta = 10$, $\omega_{\rm ph}=1$, $\omega_{\rm ct}=5$, $s=1$, $\lambda=0.1$.
}
\end{figure}

As can be easily checked, this problem has the unique solution $\ket{10100}$ (with energy $E_0=0$) and the first excited states (with a six-fold degeneracy)
are given by $\ket{00101}, \ket{10010}, \ket{00001}, \ket{00010}, \ket{01001}, \ket{01010}$ with energies $E_1=\ldots=E_6=1$.
The first excited states have Hamming distances (i.e., number of bit-flips necessary for
transformation) to the solution of $2,2,3,3,4,4$, respectively. This already indicates the hardness of such problems. 
Simple coupling Hamiltonians such as (\ref{Ehint_ex}) that are only linear in the Pauli matrices will to first order only yield transitions between states with Hamming distance 1, 
since 
$\bra{a}\sigma_i^{x/y}\ket{b} = 0$ if $\ket{a}$ and $\ket{b}$ have Hamming distance larger than one.
Of course, this does not completely prohibit transitions between states with a larger Hamming distance, but such tunneling processes will have to
pass through energetically less favorable states and are therefore strongly suppressed.
Accordingly, one may expect that the process of thermalization is strongly hampered.
This is also observed in figure \ref{Frelaxation_ec3}, where we have assumed a reservoir temperature much 
smaller than the fundamental energy gap.
Whereas for the simple qubit system (\ref{Eham_00}) the system rapidly relaxes into the ground state, this is
very different for the example (\ref{Eham_01}).


\section{Conclusion}

We have compared different procedures of deriving master equations from microscopic models and have shown that 
by using a coarse-graining approach one always obtains master equations in Lindblad form. 
This ensures for positivity and stability of the density matrix.
In contrast, the usual Markovian approximation scheme may sometimes lead to a non-positive and
even unstable behavior, where in the latter case there is no hope of approximating the exact solution.
The coarse-grained master equations depend on a parameter -- the coarse graining timescale $\tau$.
For short coarse-graining times that are adaptively matched with the physical time, the 
solution $\RS^\tau(\tau)$ must approximate the result of the Born-approximation
by construction. For large coarse-graining times and time-independent system Hamiltonians, we reproduce 
the secular approximation.
For all intermediate coarse-graining times, a positive evolution of the system density matrix is ensured by 
the Lindblad form of the resulting differential equations governing the time evolution. 
For the special case of pure dephasing of a single qubit we reproduce the analytical result by \cite{lidar2001a} that 
$\RS^t(t) = \RS(t)$ yields the exact solution (which is of course equivalent in the weak-coupling limit to
the Born approximation).
For the simple example of exponentially decaying correlation functions we find by numerical simulation a 
surprisingly perfect agreement between solutions of the integro-differential Born equation and solutions 
of the adaptive coarse-graining approach.
Given an interest in the system density matrix at time $t$, we therefore propose to match the coarse-graining 
time with the physical time $\tau=t$.
In this case one has to calculate the Liouvillian matrix elements (\ref{Eliouvillian}) 
only once (for the desired time $t$) and then evolve the density matrix 
using Eqn. (\ref{Emaster_replacement}) until that time $t$.
In terms of computational complexity, this is more efficient than an evolving an integro-differential 
equation.
A similar advantage is given when the resulting master equations are so simple such that an analytical
solution in terms of the dampening matrix elements is possible.
If in contrast the system is very complex and one is interested in the density matrix at all times, 
this advantage is destroyed.
From a computational perspective, it would therefore be interesting to find the general map
\bea\label{Ederivative}
\dot{\RS}(t) = {\cal L}(t) \RS(t) = \lim_{\Delta t \to 0} \frac{\RS(t+\Delta t) - \RS(t)}{\Delta t}
\eea
which fulfills $\RS(t) = \RS^t(t)$. Formally, such a map can be found by inserting the 
solution $\RS^\tau(t) = e^{L^\tau t} \RS(0)$ into the derivative in Eqn. (\ref{Ederivative}) 
\bea
{\cal L}(t) &=& \lim_{\Delta t \to 0} \frac{1}{\Delta t}\left[e^{L^{t+\Delta t} \cdot (t+\Delta t)} e^{-L^t \cdot t} - 1\right]\nn
&=& \left[\frac{d}{dt} e^{L^t\cdot t}\right] e^{-L^t \cdot t}\,.
\eea
We have given various examples of simple qubit systems coupled to a bosonic reservoir
where thermalization which have demonstrated that thermalization of spin systems coupled 
to a bosonic bath may depend on a plethora of factors such as the initial state, complexity 
of the system Hamiltonian and complexity of the coupling.
Future research will consider the importance of the coarse graining approach for different scenarios.


\section*{Acknowledgments}

G. S. is indebted to G. Kiesslich and M. Vogl for fruitful discussions.

$^*$\,{\small\sf schaller@itp.physik.tu-berlin.de}


\begin{appendix}


\section{Neglect of back-action}\label{Abaneglect}

Throughout the paper, we make the following simplifying assumptions

{\em Initial factorization of the density matrix}:\\
By assuming 
\bea
\f{\rho}(0) = \f{\RS}(0) \otimes \f{\RB^0} = \RS(0) \otimes \RB^0
\eea
we implicitly demand that one can at $t=0$ prepare the system in a product state, 
which requires sufficient experimental control.

{\em Born approximation}:\\
For a bath much larger than the system and weak coupling it is reasonable to assume that the
back-action of the system onto the bath is small, such that the bath part of the density matrix
is hardly changed from its initial value, i.e.
\bea
\f{\rho}(t') = \f{\RS}(t') \otimes \f{\RB^0} + \ord\{\lambda\}\,.
\eea
Clearly, inserting this approximation in the third term in (\ref{Erhoexact}) is 
consistent up to second order in $\lambda$.

{\em Reservoir in Stationary State}:\\
We will assume that the reservoir is in a stationary state of the bath Hamiltonian, which implies 
$\left[\f{\RB^0}, \HB\right]=0$. One possibility of such a stationary state is to 
assume the thermal equilibrium state ($\beta^{-1} = k_{\rm B} T$ at temperature $T$)
\bea
\f{\RB^0} = \RB^0 = \frac{e^{-\beta \HB}}{\traceB{ e^{-\beta \HB}}}\,.
\eea


\section{Markov Approximation}\label{Amarkov}

Starting from the Born Eqn. (\ref{Eborn}), the Markovian approximation in \cite{breuer2002} is 
performed by replacing (in the interaction picture) \mbox{$\f{\RS}(t')\to\f{\RS}(t)$} under the integral, 
substituting \mbox{$\tau=t-t'$} and extending the integration to infinity. 
This is usually motivated by fastly decaying reservoir correlation functions (\ref{Ecorrfunc}).
By doing so, we obtain the BM Master Equation
\bea\label{Erho2}
\f{\dot{\RS}} &=& -i \traceB{\left[\f{\HI}(t), \f{\RS}(0) \f{\RB^0}\right]}\nn
&&-\int\limits_0^\infty  \traceB{\left[\f{\HI}(t), \left[\f{\HI}(t-\tau), \f{\RS}(t) \f{\RB^0}\right]\right]} d\tau\nn
&&+ \ord\{\lambda^3\}\,,
\eea
where one can now insert the decomposition (\ref{Eintham}) of the interaction Hamiltonian to obtain
\bea\label{Erho3}
\f{\dot{\RS}} &=& -i \lambda \sum_{\cal A} \expval{B_{\cal A}} \left[\f{A_{\cal A}}(t), \f{\RS}(0)\right]\nn
&&+\lambda^2 \Big\{\sum_{\cal AB} \int\limits_0^\infty \left[\f{A_{\cal B}}(t-\tau) \f{\RS}(t), \f{A_{\cal A}}(t)\right] C_{\cal AB}(\tau) d\tau\nn
&&+ {\rm h.c.}\Big\} + \ord\left\{\lambda^3\right\}\,,
\eea
with the reservoir correlation functions (\ref{Ecorrfunc}).
We will use $\expval{B_{\cal A}}=0$ throughout, since this case can 
always be generated with a suitable transformation \cite{trafo}. 
In order to evaluate the time-dependence of the operators in (\ref{Erho3}), it is useful to expand them
into eigenoperators of the system Hamiltonian (assuming $\HS$ to be time-independent)
\bea\label{Eeigop}	
A_{\cal A} &=& \sum_\omega A_{\cal A}(\omega) = \sum_\omega A_{\cal A}^\dagger(\omega) = A_{\cal A}^\dagger\,,\nn
A_{\cal A}(\omega) &=& \sum_{ab} \delta_{(E_b - E_a), \omega}
\ket{a} \bra{a} A_{\cal A} \ket{b} \bra{b}\,,
\eea
where the variable $\omega$ ranges over all energy differences of $\HS$, $\HS\ket{a}=E_a\ket{a}$ and $\delta$ is a Kronecker symbol.
These operators $A_{\cal A}(\omega)$ have the advantageous properties
\bea
\left[ \HS, A_{\cal A}(\omega)\right] &=& -\omega A_{\cal A}(\omega)\,,\nn
\left[ \HS, A_{\cal A}^\dagger(\omega)\right] &=& +\omega A_{\cal A}^\dagger(\omega)\,,\nn
\left[ \HS, A_{\cal A}^\dagger (\omega) A_{\cal B}(\omega)\right] &=& 0\,,
\eea
which implies in the interaction picture
\bea
\f{A_{\cal A}}(t) = \sum_{\omega} e^{-i \omega t} A_{\cal A}(\omega) = \sum_{\omega} e^{+i\omega t} A_{\cal A}^\dagger(\omega)\,.
\eea
With inserting the half-sided Fourier transform of the reservoir correlation functions (\ref{Eft_half})
we obtain for the master equation in (\ref{Erho3}) with $\expval{B_{\cal A}}=0$
\bea\label{Erho4}
\f{\dot{\RS}} &\approx& \lambda^2 \Big\{\sum_{\cal AB} \sum_{\omega, \omega'} e^{i(\omega'-\omega)t} \Gamma_{\cal AB}(\omega) \left[A_{\cal B}(\omega) \f{\RS}(t), A_{\cal A}^\dagger(\omega')\right]\nn
&&+ {\rm h.c.}\Big\}\,.
\eea
Finally, we observe that by re-inserting the definition of the eigenoperators (\ref{Eeigop}) in (\ref{Erho4}) and switching
back to the Schr\"odinger picture the time-dependent phase factors vanish and one obtains the
Born-Markov master Eqn. (\ref{Emaster_mknrwa}).


\section{Secular approximation}\label{Asecular}

For large times, the terms with an oscillating prefactor in Eqn. (\ref{Erho4})
will average out and by inserting
\bea
e^{i(\omega' - \omega) t} \f{\RS}(t) \approx \delta_{\omega,\omega'} \f{\RS}(t)
\eea
in Eqn. (\ref{Erho4}) and switching to the Schr\"odinger picture 
(compare also e.g. chapter 3.3 in \cite{breuer2002}) we obtain
\bea\label{Emaster_mkrwa}
\dot{\RS} &=& -i \left[\HS, \RS(t)\right]\nn
&&+\lambda^2\sum_{\omega}\sum_{\cal AB} \left\{\Gamma_{\cal AB}(\omega)
\left[A_{\cal B}(\omega) \RS, A_{\cal A}^\dagger (\omega)\right] + {\rm h.c.}\right\}\,.\nn
\eea
The advantage of the secular approximation is that we can now combine the half sided Fourier transforms to 
full (even and odd) Fourier transforms (\ref{Eevenoddft})
of the reservoir correlation functions. Inserting these definitions into (\ref{Emaster_mkrwa}), one obtains a 
Lindblad \cite{lindblad1976a} form 
\bea\label{Elindblad_rwa}
\dot{\RS} &=& -i \left[\HS, \RS(t)\right]\nn
&&-i \left[\frac{\lambda^2}{2i}\sum_{\omega} \sum_{\cal AB} \sigma_{\cal AB}(\omega) A_{\cal A}^\dagger(\omega) A_{\cal B}(\omega), \RS(t)\right]\nn
&&+\lambda^2\sum_{\omega}\sum_{\cal AB} \gamma_{\cal AB}(\omega)\times\nn
&&\times\left[
A_{\cal B}(\omega) \RS A_{\cal A}^\dagger (\omega)
- \frac{1}{2}  \left\{A_{\cal A}^\dagger(\omega) A_{\cal B}(\omega), \RS(t)\right\}\right]\,,\nn
\eea
where the positivity of the $\gamma_{\cal AB}(\omega)$-matrix is guaranteed by Bochners theorem \cite{breuer2002,alicki1987}, 
which states that the Fourier transform of a function of positive type (as are the reservoir correlation functions) gives
rise to a positive definite matrix.
In above equation, the second commutator corresponds to the unitary action of decoherence (Lamb-shift).
Finally, we can insert the operator definitions (\ref{Eeigop}) in (\ref{Elindblad_rwa}) to obtain Eqn. (\ref{Elindblad_rwa1}).
Naturally, if $\left[\HS, A_{\cal A}\right]=0$ (pure dephasing), BM and BMS approximations are equivalent.


\section{Adiabatic approximation}\label{Aadiabatic}

With inserting the ansatz
\bea\label{Eansatzu}
U(t) &=& \sum_{ab} u_{ab}(t) \exp\left\{-i\int\limits_0^t E_\alpha(t') dt'\right\}\times\nn
&&\times \ket{a(t)} \bra{b(0)}\,,
\eea
where the $\ket{a(t)}$ span an instantaneous basis (chosen to be complete and orthonormal) of the 
system Hilbert space defined via $\HS(t) \ket{a(t)} = E_a(t) \ket{a(t)}$
in the evolution equation for the time evolution operator 
\mbox{$\dot{U}(t) = -i \HS(t) U(t)$}, one obtains an equation for the expansion coefficients
\bea
\dot u_{ab} + u_{ab} \braket{a}{\dot a} = - \sum_{c \neq a} u_{cb} 
e^{-i \int\limits_0^t g_{ca}(t') dt'} \braket{a}{\dot c}\nn
\eea
with the energy gap 
\bea
g_{ca}(t') = E_c(t') - E_a(t').
\eea
With introducing the Berry phase \cite{sun1988a}
\bea
\gamma_a(t) = i \int\limits_0^t \braket{a(t')}{\dot a(t')} dt'
\eea
this can also be written as
\bea\label{Eadiabat}
\frac{d}{dt} \left(u_{ab} e^{-i \gamma_a(t)}\right) &=& - \sum_{c \neq a} u_{cb} 
e^{-i \gamma_a(t)-i \int\limits_0^t g_{ca}(t') dt'}\braket{a}{\dot c}\,,\nonumber
\eea
which gives the general time evolution of the expansion coefficients $u_{ab}$ for
any (also non-adiabatic) system Hamiltonian $\HS(t)$ if one uses the initial condition
$u_{ab}(0) = \delta_{ab}$.
The full adiabatic approximation essentially consists in setting the right hand side of above
equation to zero: For slowly varying system Hamiltonians, the change of the eigenvectors will be 
negligible such that $\braket{a}{\dot c} \approx 0$.
Note however, that in the vicinity of avoided crossings, the condition of adiabaticity relates
the maximum speed of the time evolution with the spectral properties of the system Hamiltonian
\bea
\braket{a}{\dot c} = \frac{\bra{a} \dot{\HS} \ket{c}}{E_c - E_a}\,,
\eea
see also e.g. \cite{jansen2007a,schaller2006b}.
If $\braket{a}{\dot{c}}\approx 0$, 
one obtains with the adiabatic approximation
$u_{ab}^{\rm ad}(t) = \delta_{ab} e^{i \gamma_a(t)}$
which implies for the time evolution operator in the adiabatic limit
\bea\label{Eprod_ad}
U(t) &\approx& \sum_a e^{i \gamma_a(t) -i \int\limits_{0}^t E_a(\tau) d\tau} \ket{a(t)} \bra{a(0)}\,.
\eea
Therefore, we obtain for the time-averaged operator in (\ref{Eoptimeav})
\bea
\opav{\f{\tilde A_{\cal A}}(\omega)}{[t,t+\tau]} &\approx& \sum_{ab} \ket{a(0)}\bra{b(0)}\times\nn
&&\times \frac{1}{\tau} \int\limits_t^{t+\tau} \bra{a(t')} A_{\cal A} \ket{b(t')}\times\nn
&&\times e^{-i \left[ \gamma_{ab}(t') - \omega t' - \int\limits_0^{t'} g_{ab}(\tau) d\tau\right]}dt'\nn
\eea
with $\gamma_{ab}(t')=\gamma_a(t') - \gamma_b(t')$ and $g_{ab}(\tau) = E_a(\tau) - E_b(\tau)$.


\section{Positivity-preserving master equations}\label{Apositivity}

Here we will show that master equations of the form
\bea\label{Epseudolindblad}
\dot{\rho} &=& -i \left[ {\cal H}(t), \rho(t)\right]\nn
&&+\sum_{\alpha\beta=1}^K \Gamma_{\alpha\beta}(t) \Big[
{\cal L}_\alpha(t) \rho(t) {\cal L}_\beta^\dagger(t)\nn
&&- \frac{1}{2} \left\{{\cal L}_\beta^\dagger(t) {\cal L}_\alpha(t), \rho(t)\right\}\Big]
\eea
generally preserve the positive semidefiniteness of an initial condition $\rho(0)$ if the
matrix $\Gamma_{\alpha\beta}(t)$ is 
positive semidefinite 
and the operator ${\cal H}(t)={\cal H}^\dagger(t)$ is hermitian at all times (smoothness of all time-dependencies
provided).
With discretizing the time derivative and by introducing new operators
\bea
W_1(t) &=& \f{1} = W_1^\dagger(t)\,,\nn
W_2(t) &=& i {\cal H}(t) + \frac{1}{2} \sum_{\alpha\beta} \Gamma_{\alpha\beta}(t) {\cal L}_\beta^\dagger(t) {\cal L}_\alpha(t)\,,\nn
W_3(t) &=& {\cal L}_1(t)\,,\nn
&\vdots&\nn
W_{K+2}(t) &=& {\cal L}_K(t)
\eea
we obtain
\bea\label{Eposmap}
\rho(t+\Delta t) = \sum_{\alpha\beta=1}^{K+2} w_{\alpha\beta}(t) W_\alpha(t) \rho(t) W_\beta^\dagger(t)
\eea
where the $w_{\alpha\beta}(t)$ matrix is given by
\bea
w(t) = \left(
\begin{array}{cc|c}
1 & -\Delta t & 0 \cdots 0\\
-\Delta t & 0 & 0 \cdots 0\\
\hline
0 & 0 & \\
\vdots & \vdots & \Delta t \Gamma(t)\\
0 & 0
\end{array}
\right).
\eea
This matrix has a simple block structure and it is therefore straightforward to relate the eigenvalues of $w$ to 
those of $\Gamma$. In particular, one obtains the eigenvalues $w_1 = 1/2\left[1-\sqrt{1+4\Delta t^2}\right]$, $w_2 = 1/2\left[1+\sqrt{1+4\Delta t^2}\right]$, 
and $w_{i\ge3}(t) = \Delta t \gamma_{i-2}(t)$, 
where $\gamma_{i-2}(t)$ are the non-negative eigenvalues of $\Gamma(t)$.
With diagonalizing the matrix in (\ref{Eposmap}) via $w_{\alpha\beta}(t) =\sum_{\gamma} u_{\alpha\gamma}(t) u_{\beta \gamma}^*(t) w_\gamma(t)$ with
a suitable (time-dependent) unitary transformation $U(t)$ we obtain from Eqn. (\ref{Eposmap})
\bea
\rho(t+\Delta t) &=& \sum_{\gamma} w_\gamma(t) \tilde W_\gamma(t) \rho(t) \tilde W_\gamma^\dagger(t)\,,\nn
\tilde W_\gamma(t) &=& \sum_\alpha u_{\alpha\gamma}(t) W_\alpha(t)\,.
\eea
Assuming that at time $t$ we have a valid density matrix with $0 \le \rho_\delta(t) \le 1$ we obtain by inserting the spectral 
decomposition $\rho(t) = \sum_\delta \rho_\delta(t) \ket{\Phi_\delta(t)}\bra{\Phi_\delta(t)}$ 
\bea
\bra{\Phi} \rho(t+\Delta t) \ket{\Phi} &=& \sum_{\gamma\delta} w_\gamma(t) \rho_\delta(t) \abs{\bra{\Phi} \tilde W_\gamma \ket{\Phi_\delta(t)}}^2\nn
&\ge& \frac{1}{2}\left[1-\sqrt{1+4\Delta t^2}\right] \times\nn
&&\times\sum_\delta \rho_\delta(t) \abs{\bra{\Phi} \tilde W_1 \ket{\Phi_\delta(t)}}^2\nn
&\ge& -\Delta t^2 \sum_\delta \rho_\delta(t) \abs{\bra{\Phi} \tilde W_1 \ket{\Phi_\delta(t)}}^2\nn
&\stackrel{\Delta t \to 0}{\ge}& 0\,,
\eea
such that in the limit $\Delta t \to 0$ (which defines the original differential Eqn. (\ref{Epseudolindblad})), the 
smallest eigenvalue of the density matrix at time $t+\Delta t$ approaches zero faster than the discretization width $\Delta t$.
Therefore, in this limit the matrix $w(t)$ becomes positive semidefinite and the differential Eqn. (\ref{Eposmap}) 
becomes a positivity-preserving map.


\section{Sinc Distribution}\label{Asincidentity}

For discrete $a,b$ and continuous $\omega$ we would like to analyze
\bea\label{Efunction}
f(\omega, a, b) &\equiv& \lim_{\tau\to\infty} \tau \; \sinc{(\omega+a)\frac{\tau}{2}} \sinc{(\omega+b)\frac{\tau}{2}}\nn
&=& \lim_{\tau\to\infty} \frac{4 \sin\left[ (\omega+a)\frac{\tau}{2}\right] \sin\left[(\omega+b)\frac{\tau}{2}\right]}{(\omega+a)(\omega+b)\tau}\,.
\eea
One can consider the case $a\neq b$ by a partial fraction expansion
where without loss of generality we find for the first term (due to the symmetry the second term can be treated in a completely analogous way)
\bea
I_{a\neq b} &=& \int\limits_{-\infty}^{+\infty} g(\omega) \frac{\sin\left[ (\omega+a)\frac{\tau}{2}\right] \sin\left[(\omega+b)\frac{\tau}{2}\right]}{\tau (\omega + a)} d\omega\nn
&=& \cos\left[(b-a)\frac{\tau}{2}\right] \int\limits_{-\infty}^{+\infty} g(\omega)\frac{\sin^2\left[ (\omega+a)\frac{\tau}{2}\right]}{\tau(\omega+a)} d\omega\nn
&&+\frac{1}{2}\sin\left[(b-a)\frac{\tau}{2}\right] \int\limits_{-\infty}^{+\infty} g(\omega)\frac{\sin\left[ (\omega+a) \tau\right]}{\tau(\omega+a)} d\omega\,,\nonumber
\eea
where we have inserted $(\omega+b) = (\omega+a) + (b-a)$ to use
trigonometric relations for $\sin(\alpha+\beta)$. With a suitable transformation, this becomes
\bea
I_{a\neq b} &=& \frac{\cos\left[(b-a)\frac{\tau}{2}\right]}{\tau} \int\limits_{-\infty}^{+\infty} g\left(\frac{x}{\tau}-a\right) \frac{\sin^2(x/2)}{x} dx\nn
&&+ \frac{\sin\left[(b-a)\frac{\tau}{2}\right]}{2 \tau}  \int\limits_{-\infty}^{+\infty} g\left(\frac{x}{\tau}-a\right) \frac{\sin(x)}{x} dx\,,
\eea
where for large $\tau$, the dominant integral contribution evaluates the (smooth) function $g(\omega)$ near $g(-a)$ such that this value
can be taken out of the integral and we obtain 
$\lim\limits_{\tau\to\infty} I_{a\neq b}=0$.
Using the representation
\bea\label{Esincdelta}
\lim\limits_{\tau\to\infty} \tau\, {\rm sinc}^2\left[\Omega\tau\right] = \pi \delta(\Omega)
\eea
one can consider the general case via
\bea
f(\omega, a, b) &\asymp&\delta_{ab} \lim_{\tau\to\infty}  \tau \; {\rm sinc}^2\left[(w+a)\frac{\tau}{2}\right]\nn
&=&\delta_{ab} \pi \delta\left(\frac{w+a}{2}\right)\,,
\eea
which yields Eqn. (\ref{Esincidentity}).

\end{appendix}



\begin{thebibliography}{9999}
\bibitem{trafo}
The general case $\expval{B_{\cal A}}\neq 0$ can be reduced to the case $\expval{B'_{\cal A}}=0$ by using the
transformations 
\mbox{$\HS\to\HS'=\HS+\lambda g_{\cal A}\sum_{\cal A} A_{\cal A}$} and
\mbox{$B_{\cal A}\to B_{\cal A}'=B_{\cal A}- g_{\cal A}\f{1}$} 
in the Schr\"odinger picture, where the real constant has to be adapted such that 
$g_{\cal A} = \expval{B_{\cal A}}$.
\end{thebibliography}
\end{document}